\documentclass[conference]{IEEEtran}

 \usepackage{fancyhdr} 
\ifCLASSINFOpdf
  \usepackage[pdftex]{graphicx}
\else
  \usepackage[dvips]{graphicx}
\fi
\usepackage[cmex10]{amsmath}
\usepackage{url}
\usepackage{listings}
\usepackage{placeins}
\usepackage{fancyvrb}
\usepackage{cite}

\usepackage{algorithm}
\usepackage[noend]{algpseudocode}
\algrenewcommand\algorithmicindent{0.15em}%
\makeatletter
\algrenewcommand\ALG@beginalgorithmic{\small}
\algrenewcommand\algorithmiccomment[2][\normalsize]{{#1\hfill\(\triangleright\) #2}}
\makeatother

\usepackage{graphicx}
\usepackage{caption}
\usepackage{subcaption}
\usepackage{tabularx}
\usepackage{enumitem}
\graphicspath{ {./figures/} }
\renewcommand{\footnotesize}{\scriptsize}
\makeatletter
\algrenewcommand\ALG@beginalgorithmic{\footnotesize}
\def\algbackskip{\hskip\dimexpr-\algorithmicindent-\labelsep}
\def\LState{\Statex \algbackskip}
\makeatother

\usepackage{algorithm}
\usepackage[noend]{algpseudocode}
\algnewcommand{\LineComment}[1]{\State \emph{\textcolor{blue}{\(\triangleright\) #1}}}
\algrenewcommand\algorithmicindent{1em}%

\usepackage{color, soul} 
\soulregister\emph7
\soulregister\cite7
\soulregister\ref7

\begin{document}
\setlength{\abovedisplayskip}{3pt}
\setlength{\belowdisplayskip}{3pt}

\setlength\floatsep{0.3\baselineskip plus 1pt minus 20pt}
\setlength\textfloatsep{0.3\baselineskip plus 1pt minus 20pt}
\setlength\intextsep{0.3\baselineskip plus 1pt minus 20pt}
%\IEEEoverridecommandlockouts
%\IEEEpubid{
%\parbox[t]{\columnwidth}{
%SC16; Salt Lake City, Utah, USA; November 2016 \\
%978-1-4673-8815-3/16/\$31.00 \copyright 2016 IEEE \hfill} \hspace{\columnsep}\makebox[\columnwidth]{}
%}

\title{Optimizing Deep Learning Recommender Systems' Training On CPU Cluster Architectures}



\author{\IEEEauthorblockN{Dhiraj Kalamkar\IEEEauthorrefmark{1}, 
Evangelos Georganas\IEEEauthorrefmark{2},
Sudarshan Srinivasan\IEEEauthorrefmark{1},\\
Jianping Chen\IEEEauthorrefmark{3},
Mikhail Shiryaev\IEEEauthorrefmark{4}, and
Alexander Heinecke\IEEEauthorrefmark{2}
}
\IEEEauthorblockA{\IEEEauthorrefmark{1}Intel Technology India Private Ltd., India}
\IEEEauthorblockA{\IEEEauthorrefmark{2}Intel Corporation, USA}
\IEEEauthorblockA{\IEEEauthorrefmark{3}Intel China Co. Ltd., China}
\IEEEauthorblockA{\IEEEauthorrefmark{4}Intel Russia, Russia}
}

%\author{\IEEEauthorblockN{Evangelos Georganas\IEEEauthorrefmark{1},
%Hans Pabast\IEEEauthorrefmark{2},
%Greg Henry\IEEEauthorrefmark{3}, and
%Alexander Heinecke\IEEEauthorrefmark{1}
%}
%\IEEEauthorblockA{\IEEEauthorrefmark{1}Intel Corporation,
%2200 Mission College Blvd.,
%Santa Clara, 95054, CA, USA}
%\IEEEauthorblockA{\IEEEauthorrefmark{2}Intel Semiconductor AG,
%Badenerstrasse 549,
%8048 Zurich, Switzerland}
%\IEEEauthorblockA{\IEEEauthorrefmark{3}Intel Corporation,
%2111 NE 25$^{\text{th}}$ Avenue,
%Hillsboro, 97124, OR, USA}
%}

\maketitle
%\thispagestyle{fancy}
%\lhead{}
%\rhead{}
%\chead{}
%\lfoot{\footnotesize{
%SC20, November 15-20, 2020, Atlanta, Georgia, USA 
%\newline 978-1-5386-8384-2/18/\$31.00 \copyright 2018 IEEE
%}} \rfoot{} \cfoot{} \renewcommand{\headrulewidth}{0pt}
%\renewcommand{\footrulewidth}{0pt}

\begin{abstract}
During the last two years, the goal of many researchers has been to squeeze the last bit of performance out of HPC system for AI tasks. Often this discussion is held in the context of how fast ResNet50 can be trained. Unfortunately, ResNet50 is no longer a representative workload in 2020. Thus, we focus on Recommender Systems which account for most of the AI cycles in cloud computing centers. More specifically, we focus on Facebook's DLRM benchmark. By enabling it to run on latest CPU hardware and software tailored for HPC, we are able to achieve more than two-orders of magnitude improvement in performance (110x) on a single socket compared to the reference CPU implementation, and high scaling efficiency up to 64 sockets, while fitting ultra-large datasets. This paper discusses the optimization techniques for the various operators in DLRM and which component of the systems are stressed by these different operators. The presented techniques are applicable to a broader set of DL workloads that pose the same scaling challenges/characteristics as DLRM.
\end{abstract}

% \fi
\IEEEpeerreviewmaketitle

\section{Introduction and Background}
\label{sec:intro}

Over the last two years, there has been steep increase in the number of Bird of a Feather (BoFs) sessions or workshops on high performance deep learning at major
supercomputing venues, such as Supercomputing (SC) and International
Supercomupting Conference (ISC). A similar trend can be seen 
in technical paper tracks focusing on Machine Learning at similar venues. While the presented
research is focusing on the extreme-scale and high performance computing (HPC)
aspect, it is often limited to training convolutional neural nets (CNN).
In essence, researchers study and explore how ResNet50 can be trained on thousands
of (accelerated) nodes in less than a minute~ \cite{Dryden:2019:CFP:3295500.3356207,DBLP:journals/corr/abs-1711-04325,DBLP:journals/corr/abs-1709-05011,DBLP:journals/corr/abs-1811-05233}. In contrast to this, the "Super 7", namely Facebook, Google, Microsoft, Amazon, Baidu, Alibaba, and Tencent, have 
published that the CNNs are only contributing a single digit or very low double
digit percentage to their workload mix~\cite{DBLP:journals/corr/JouppiYPPABBBBB17,Zion2019,ArchImpl19}. Thus, we believe, the HPC research community needs to shift
its focus away from CNN models to the models which have the highest percentage
in the Super 7's application mix: 1. recommender systems (RecSys) and 
2. language models, e.g.\ recurrent
neural networks/long short term memory (RNN/LSTM), and attention/transformer.
For the second category, some research has been already published on how to 
scale it to large platforms  \cite{8891019,You:2019:LTL:3295500.3356137}, whereas RecSys's detailed analyses
are basically non-existent in the supercomputing context. This is mainly due to the fact that the industry/the Super 7
have not aligned so far on a standard benchmark (such as ResNet50 for CNNs) for these
types of neural networks. Therefore the current release of MLPerf Training \cite{Mattson2019MLPerfTB} is only
using a small NCF model. These models do not captures real-life behavior correctly, in terms of model-design and size of the model.

To address these concerns, Facebook recently proposed a deep learning recommendation model (DLRM)~\cite{Naumov2019DeepLR}. Its purpose is to allow
hardware vendors and cloud service providers to study different system configurations, or in simple
words to allow for a systematic hardware-software co-design exploration for deep learning systems
DLRM comprises of the following major components:
a) a sparse embedding realized by tables (databases) of various sizes, and b) small dense multi-layer perceptron (MLP). Both a) and b)
interact and feed into c) a larger and deeper MLP. Note that all three parts can be 
configured (number of features, mini-batch sizes and table sizes) to re-balance and study 
different topology configurations in a straightforward manner. 
This work unveils that DLRMs mark the start of a new era of deep learning workloads. In contrast
to CNNs, RNNs or Transformers they stress all properties and components of a computing
system. This is due to the sparse model portion and as well the dense model portion and
of course the communication and interaction between these two. While the sparse portion
challenges the memory capacity and bandwidth side, and the dense portion the compute capabilities
of the system, the interaction stresses the interconnect. We therefore recognize that a system
which maximizes DLRM training performance needs a balanced design between memory capacity, memory bandwidth, interconnect bandwidth and compute/floating point performance. 

Facebook open-sourced a simple single process reference implementation in PyTorch and Caffe2\cite{dlrm2019}. This implementation yields good performance numbers on GPUs (due to PyTorch GPU-affinity), but CPUs are still a 2\textsuperscript{nd} class citizen in PyTorch. 
Due to DLRMs' large memory footprint for scale-out
scenarios (e.g. real-life terabytes of datasets), we focus on CPUs in this work to understand the scaling implications. On GPUs we would
have to scale immediately to very high GPU counts (potentially even spanning embedding tables across multiple GPUs) due to the very limited capacity of HBM/GDDR per GPU.
The problem sizes we will tackle, will not fit onto Nvidia DGX boxes in most of the cases.
Additionally, at least up to now, it has been challenging to converge DLRM with FP16 using the default optimizers, such as stochastic gradient descent.
This is due to the limited accuracy (not enough mantissa bits) of FP16~\cite{Zhang2018TrainingWL}. However, a novel scheme that utilizes the BFLOAT16  datatype
is introduced in this work. It demonstrates convergence to state-of-the-art accuracies while using DLRM's default optimizer. Therefore, one of the major selling
point of GPU's for deep learning, the FP16 tensor cores, cannot be easily leveraged in DLRM with its default optimizers. A comparison between CPU and GPU using FP32 will be covered for a smaller dataset on a single-socket/single GPU basis. 

Our contributions are:
\begin{itemize}
    \item We present a bottom-up performance analysis of standalone micro-apps to identify
    even more optimization/performance potential in the DLRM code if being freed from DL framework restrictions, and to inform future PyTorch/framework enhancements
    \item We present a top-down performance analysis and PyTorch code changes that speed-up the reference DLRM topology by roughly two orders of magnitude (110$\times$) on single socket latest Intel Xeon processors. The changes and improvements are upstreamed to PyTorch as part of an active 
    pull-request on github at the time of this writing
    \item In the DLRM code, we enable hybrid parallelism (running the sparse embedding workload portion in a model-parallel and the MLPs in data-parallel fashion) for multi-process parallelization, targeting multi-node Xeon experiments with optimized communication libraries
    \item We discuss and analyze the scaling properties of our code on a fat shared-memory system 
    (octo-socket Intel Xeon with 224 cores, to emulate Facebook's recently announced Zion training platform \cite{Zion2019}) and traditional HPC clusters (dual-socket Intel Xeon
    with 56 cores and HPC interconnect, Intel Omnipath (OPA)). We present
    strong and weak-scaling experiments for various datasets, including the benchmark proposed to MLPerf~\cite{Mattson2019MLPerfTB}.
    \item Last but not least, we introduce a novel implementation of the stochastic gradient (SGD) optimizer targeting mix-precision training, called Split-SGD-BF16. It avoids the need for master weights in case of BFLOAT16 mixed precision 
    training. Master weights are nearly always needed 
    in case of FP16 training which complicates the software development. This 
    new SGD future-proofs our work toward next-generation CPUs and accelerators, and is transferable to all other deep learning topologies such as CNNs, RNN/LSTMs, etc.
\end{itemize}

These contributions are discussed in the following structure: We start with summarizing 
the DLRM architecture in Sect.~\ref{sec:dlrm}. In terms of covering the mentioned optimization,
we split our work into single-socket, multi-socket implementation and performance analysis
(Sect.~\ref{sec:singlesocket}) through (Sect.~\ref{sec:results}). Before concluding our research, we will introduce the novel Split-SDG-BF16
as an outlook to next-generation CPUs in Sect.~\ref{sec:splitsgd}. Finally in Sect.~\ref{sec:relwork} and Sect.~\ref{sec:conclusion} we conclude
our work.
\section{The DLRM Benchmark}
\label{sec:dlrm}

\begin{figure}[!t]
  \begin{center}
  \includegraphics[width=1.0\linewidth]{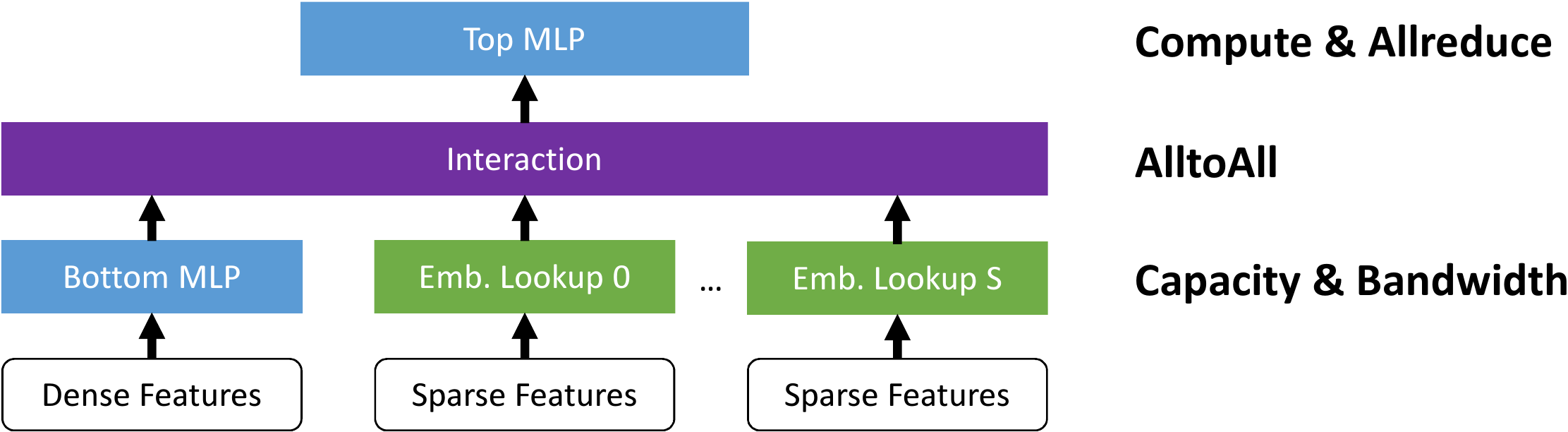}
  \end{center}
  \caption{Schematic of the DLRM topology. It comprises of MLPs and Embedding table look-ups 
  and the corresponding interaction operations. Thus, it stress all important aspects of the underlying hardware platform at the same time: 
  compute capabilities, network bandwidth, memory capacity and memory
  bandwidth. This is rather unusual for classic HPC applications.}
  \label{fig:dlrm}
\end{figure}

In this section, we briefly summarize the DLRM benchmark as proposed and published
by Facebook~\cite{Naumov2019DeepLR,Zion2019}. As already touched upon in the introduction, 
the DLRM benchmark aims at capturing Facebook's most important deep learning training workload 
through a customize-able benchmark that can be shared with the research community outside
of Facebook. 

Figure~\ref{fig:dlrm} depicts a bird's eye-view of the application. DLRM combines sparse and dense
features, which are normally provide through the network in case of online inference, to provide
a recommendation for these inputs, e.g. to recommend a specific product for a specific user.
The dense features are piped through a classic neural network which is in this case a model problem
of a multi-layer perceptron identified as Bottom MLP. MLPs are a very classic concept of neural nets
and when using them in a minibatched fashion they transfer computationally into a sequence of matrix
multiplications (GEMM) followed by an activation function calculation such as ReLU or Sigmoid. As
the activation function calculation is element-wise, it is complexity-wise irrelevant. This can 
be also seen when examining runtime statistics. As the Bottom MLP is normally relatively small it
is somewhat compute bound due to its GEMM-nature but also demands relatively high memory bandwidth  as the GEMMs 
are small and have little re-use. In terms of multi-socket parallelization, MLPs are parallelized best
using data-parallelism. This strategy leads to splitting the minibatch across different sockets
and requires an allreduce of the weight-gradients during backward propagation, cf. \cite{Dryden:2019:CFP:3295500.3356207,DBLP:journals/corr/abs-1711-04325,DBLP:journals/corr/abs-1709-05011,DBLP:journals/corr/abs-1811-05233}. 

The sparse features, which handle categorical
data, are mapped through embeddings into a dense space such that they can be combined with the output
of the Bottom MLP. We can think of embedding as mulit-hot encoded look-ups into an embedding table $W \in \mathcal{R}^{MxE}$, with $M$ denoting the entries and $E$ the length of each entry. That means, the multi-hot weight-vector $\alpha^T = [0,\ldots,a_{p_{1}},\ldots,a_{p_{k}},\ldots,0]$ with elements $a_p \neq 0$ for $p=p_1, \ldots, p_k$ and $0$
elsewhere with $p$ being the index for the corresponding lookup items. When adding the minibatch $N$,
we can rewrite the embedding look-up as $L = A^T W$ with $L \in \mathcal{R}^{NxE}$ and the 
sparse matrix $A = [a_1, \ldots, a_N]$. For more details we refer to~\cite{DBLP:journals/corr/abs-1901-02103}. We further want to note that full rows are read
from $W$, thus frameworks provide special dense vector routines for embedding lookups and they do not 
rely on sparse matrix multiplication for its implementation. As shown in Fig.~\ref{fig:dlrm}, there
are normally many dense embedding tables and they can have millions of entries. Therefore the 
combined task's performance is dictated by available memory capacity and bandwidth. 

The results of the Bottom MLP and the embedding look-ups are combined in a so-called Interaction which produces a condensed output. Concat is an example of simple interaction operation. More commonly a self dot product is used as an interaction op which translates to a batched matrix-matrix multiplication as a key kernel. As we can see this operation is parallelized best over the different embedding tables. Therefore, a personalized all-to-all communication is needed within the interaction 
operation to switch from model parallel to data parallel execution.
%This can be seen as an \texttt{MPI\_Alltoall} in terms of dataflow and batched-matrix multiplication. Especially the all-to-all communication after the embedding can be hidden while running the Bottom MLP, turning the interaction into an aspect of non-concern. 
Finally, the condensed data is fed into a wider and deeper MLP, the Top MLP. Most of the dense 
floating operations of the execution are spent here since it is merely a GEMM. As in case of the Bottom MLP,
this MLP is parallelized best using data-parallel execution.

For computing the loss, DLRM utilizes the classic cross-entropy loss function which is not further
discussed as it does not result into any performance implications at all. Therefore, it is not 
even depicted in Fig.~\ref{fig:dlrm}.

In terms of computation, we can summarize that the following patterns require attention including
concrete performance expectations:
\begin{itemize}
    \item \textbf{GEMM:} the sizes are often (especially in case of scaling out) non-squared, so we need the best possible GEMM routines, and data layouts in order to reach the compute-bound limit of the machine. 
    \item \textbf{Embedding Look-ups:} this is a GUPS-like~\cite{5161019} kernel, however we read several consecutive cachelines from memory and sum them up. The expectation is that these operations run at close 
    to peak bandwidth performance of the machine.
    \item \textbf{Allreduce:} for reducing the weigth gradients in the backward pass of the MLPs we 
    need to use the best allreduce algorithm and ensure overlap with the GEMM compute.
    \item \textbf{Alltoall:} for switching between data and model parallelism during the interaction operation
\end{itemize}
All other kernels and primitives such as activation functions and loss computations are 
not covered
in this work in detail as they do not take a huge fraction of the time and can be relatively easily overlapped 
or fused with other operations, e.g. ReLU can directly happen inside a custom GEMM routine when the C
matrix is still hot in caches~\cite{Georganas:2018:AHD:3291656.3291744}.

This description points out clearly that DLRMs stress all 
components of the (cluster) system: compute and network,
memory bandwidth and network, capacity and compute. Typical HPC applications as well as CNN training usually stress only one or two
aspects. More specifically, DLRM feels like running HPCG~\cite{10.1093/nsr/nwv084} and HPL~\cite{hplpaper} in a single and tightly coupled benchmark. This mix of computational motifs poses yet another challenge in designing balanced AI-oriented systems and impacts the machine of choice. Therefore, we will also later study 
unconventional hardware platforms such as 8 socket servers. 
\section{Implementation and Optimizations - Single Socket}
\label{sec:singlesocket}

In this section we present characteristics of key compute kernels in DLRM, potential issues with their efficient implementation 
and our bottom up approach for evaluating different alternatives. We then try to integrate these back into PyTorch framework wherever possible 
to evaluate end-to-end performance gains. In few cases, we just show the estimated benefits of our optimizations if made free from framework restrictions.

\subsection{Bottom-Up - Embedding Look up Layer}
\begin{algorithm}[t]
\begin{algorithmic}[1]
\LState \textbf{Inputs}: Weight $W[M][E] $, Indices $I[N+1]$, Offsets $O[NS]$
\LState \textbf{Outputs}: Output $Y[N][E]$
\For{$n = 0 \dots N$}
\State $start = O[n]$
\State $end = O[n+1]$
\State $Y[n][:] = 0.0$
\For{$s = start \dots end$}
\State $ind = I[s]$
\For{$e = 0 \dots E$}
\State $Y[n][e] \mathrel{+}= W[ind][e]$
\EndFor
\EndFor
\EndFor
\end{algorithmic}
\caption{Forward pass of EmbeddingBag Layer}
\label{alg:embfwd}
\end{algorithm}

\begin{algorithm}[t]
\begin{algorithmic}[1]
\LState \textbf{Inputs}: GradOut $dY[N][E]$, Offsets $O[N+1]$
\LState \textbf{Outputs}: GradWeight $dW[NS][E]$
\For{$n = 0 \dots N$}
\State $start = O[n]$
\State $end = O[n+1]$
\For{$s = start \dots end$}
\For{$e = 0 \dots E$}
\State $dW[s][e] \mathrel{+}= dY[n][e]$
\EndFor
\EndFor
\EndFor
\end{algorithmic}
\caption{Backward pass of EmbeddingBag Layer}
\label{alg:embbwd}
\end{algorithm}

\begin{algorithm}[t]
\begin{algorithmic}[1]
\LState \textbf{Inputs}: GradWeight $dW[NS][E]$, Indices $I[NS]$m $\alpha$
\LState \textbf{InOut}: Weight $W[M][E]$
\For{$i = 0 \dots NS$}
\State $ind = I[i]$
\For{$e = 0 \dots E$}
\State {Perform Atomic} 
\State \hskip1.0em $W[ind][e] \mathrel{+}= \alpha * dW[i][e]$
\EndFor
\EndFor
\end{algorithmic}
\caption{Update pass of Sparse EmbeddingBag Layer}
\label{alg:embupd}
\end{algorithm}

\begin{algorithm}[t]
\begin{algorithmic}[1]
\LState \textbf{Inputs}: GradWeight $dW[NS][E]$, Indices $I[NS]$m $\alpha$
\LState \textbf{InOut}: Weight $W[M][E]$
\State {for each thread $tid$ in $nThreads$}
\State $M\_start = (M * tid) / nThreads$
\State $M\_end = (M * (tid+1)) / nThreads$
\For{$i = 0 \dots NS$}
\State $ind = I[i]$
\If {$ind >= M\_start$ $and$ $ind < M\_end$}
\For{$e = 0 \dots E$}
\State $W[ind][e] \mathrel{+}= \alpha * dW[i][e]$
\EndFor
\EndIf
\EndFor
\end{algorithmic}
\caption{Race Condition Free Update for Sparse EmbeddingBag Layer}
\label{alg:embupd2}
\end{algorithm}

In a typical DL framework, a training iteration consist of 3 passes. A forward pass which computes loss with respect to model parameters, a backward pass which computes gradients with respect to model parameters and an update pass (typically run by an optimizer) which takes learning rate and computed gradients from backward pass and updates the model parameters.
Algorithms \ref{alg:embfwd}, \ref{alg:embbwd} and \ref{alg:embupd} shows the forward, backward and the update pass for Embedding table look up and sparse update operation in typical DLRM code when using a linear solver like SGD. The forward and backward pass can be trivially parallelized using \texttt{\#pragma omp parallel for} over the first line of both forward and backward pass. Compiler can easily vectorize these operations over E loop. The update operation is also trivially simple. However, parallelizing it over NS loop introduces a potential race condition due to indirect access on output rows. Essentially, we need atomicity while performing accumulation to avoid the race condition. Intel Xeon processors provide atomic operations only on integer data types. However, we need floating point atomic\_add. There are multiple approaches to solve this problem as discussed below:
\begin{enumerate}
    \item Floating point atomic\_add using atomic xchg
    \item Using Intel Restricted Transactional Memory (RTM)
\end{enumerate}
Even though we prefer RTM approach over atomic xchg as it allows to use SIMD vector multiply-add instructions inside critical section, we expect both of these approaches perform similarly due to their memory bandwidth limited nature when most of the indices are unique and there is little contention. However, since both of these approaches require a cache line to be modified in its own cache, if there are significant repeated use of indices, there is a potential issue due to excessive cache line thrashing across the core caches when multiple entries of same index are distributed across cores. To address this issue, we implemented the update pass using a race free algorithm and expect it to perform better  in a contentious environment/setup. The idea is shown in Algorithm \ref{alg:embupd2}. We simply divide the total rows in table among available threads, and each thread goes over the full list of indices but updates the rows only if it belongs to its own range of rows. This algorithm not only eliminates the race conditions but also helps to improve locality of accesses. However, this approach has a potential load imbalance issue if indices are clustered rather than being distributed uniformly across all the rows.

Note that in theory the backward and update pass for EmbeddingBag can be fused into
 a single operation yielding better performance. However, such fusing are not currently implemented by deep learning framework principles. Nevertheless, we did such an implementation 
in standalone that achieves up to 1.6x speed-up for embedding updates. However, due to difficulty of integration, we will not consider it any further in this work.

%\subsection{Bottom-Up - Fused Embedding Backward}

%\begin{figure}[!t]
%  \begin{center}
%  \includegraphics[width=1.0\linewidth]{figs/emb_fused.pdf}
%  \end{center}
%  \caption{Performance optimization from fusing backward and embedding table update passes using
% Intel TSX (transactional memory extensions).}
%  \label{fig:embedding_fused}
%\end{figure}

%In case of the backward embedding pass, which accounts for the highest performance improvement 
%in the previous sections, even more is possible when freeing the operation from the
%framework. PyTorch requires us to have a backward by data and optimizer steps (the actual update) in separate calls. 
%The optimizer in this case is stochastic gradient descent (SGD). However, mathematical there
%is no reason that these two backward operations cannot be fused into a single one. Of course
%there is the challenge that we have to preform now random updates to the embedding tables which 
%can collide. CPUs do not offer floating point atomics, so at first sight the only option is 
%fine grain locks. We tried using Intel TSX, the transaction memory
%support in Skylake. This feature was specifically designed for database updates and embedding tables
%can be seen as databases. Figure~\ref{fig:embedding_fused} shows that backward embedding
%time can be further halved using this approach.

\subsection{Bottom-Up - Multilayer Perceptron (MLP)}
\label{sec:mlp}

A MLP consists of (at least three) \emph{fully connected} layers of neurons: the topology starts with an input layer, followed by a number of hidden layers which conclude to the output layer. For the rest of this section we consider the optimization of the \emph{fully connected} layers since they constitute the cornerstone of MLP.
Mathematically, an input layer $x\in {\rm I\!R}^{C}$ is mapped to an output layer $y\in {\rm I\!R}^{K}$ via the relation $y=W\cdot x$, where $W\in {\rm I\!R}^{K\times C}$ is the weight tensor of the connections between the neurons. During the training process, $N$ multiple inputs ($N$ is the so-called mini-batch size) are grouped together yielding the equation $Y=W\cdot X$ with $W\in {\rm I\!R}^{K\times C}$,  $X\in {\rm I\!R}^{C\times N}$ and $Y\in {\rm I\!R}^{K\times N}$. Observe that by increasing the mini-batch $N$, we fundamentally increase the weight tensor reuse. Typical implementations of Fully Connected layers (e.g.\ the one in the PyTorch DLRM version) leverage a large GEMM call and they apply the activation functions onto the GEMM outputs. Even though such an approach is straightforward to implement, its performance can be underwhelming for two reasons: i) typical high-performance GEMM library calls internally perform packing of sub-matrices to ameliorate TLB misses and cache conflict misses~\cite{goto2008anatomy}, and ii) the multi-threaded implementation of GEMM with shapes arising from small mini-batch values $N$ may not fully exploit the available data reuse.

In this section we dive into the details of the forward propagation algorithm of the MLP training process (also used for inference); we also implemented all the required kernels of the back-propagation training in an analogous fashion. Our MLP implementation which leverages the batch-reduce GEMM microkernel follows the same principles of previous work~\cite{georganas2020ipdps}, and we present it here for completeness. However, in this work we modify the tensor layouts since we found this variation to yield better performance in training where the activations tensors in the backward by weights pass are the analogous of the weights tensors in the forward/backward by data passes. We emphasize upfront here that incorporating blocked tensor layouts in frameworks (for performance reasons) is rather cumbersome, yet we examine with standalone MLP training code the potential upside. 

\begin{algorithm}[t]
\begin{algorithmic}[1]
 \LState \textbf{Inputs}: Weight $W[K_b][C_b][b_c][b_k]$, Input $X[C_b][N_b][b_n][b_c]$
\LState \textbf{Outputs}: Output $Y[K_b][N_b][b_n][b_k]$
\State Based on $thread\_id$ calculate $K_b\_start$, $K_b\_end$, $N_b\_start$ and $N_b\_end$ to assign output work items
\For{$ib_n=N_b\_start \dots N_b\_end$}
\For{$ib_k=K_b\_start \dots K_b\_end$}
\State{/* Prepare batch-reduce GEMM arguments */}
\For{$ib_c=0 \dots C_b-1$}
\State $A_{ptrs}[ib_c] = \&W[ib_k][ib_c][0][0]$
\State $B_{ptrs}[ib_c] = \&X[ib_c][ib_n][0][0]$
\EndFor
\State $Out =  \&Y[ib_k][ib_n][0][0]$
\State $\mathbf{batchreduce\_gemm}(A_{ptrs}, B_{ptrs}, Out, C_b)$ 
\EndFor
\EndFor
\end{algorithmic}
\caption{Forward pass of Fully Connected Layer}
\label{alg:mlp}
\end{algorithm}

Algorithm~\ref{alg:mlp} shows the implementation of the forward propagation in the training process of fully connected layers. First, we highlight the blocked tensor layout; all the 2-dimensional tensors are transformed into 4-dimensional ones by blocking the mini-batch dimension $N$ with a factor $b_n$ and the tensor dimensions $C$ and $K$ with blocking factors $b_c$ and $b_k$ respectively. Such a blocked layout exposes better locality and avoids large, strided sub-tensor accesses which are known to cause TLB misses and cache conflict misses in case the leading dimensions are large powers of 2. Also, in contrast to previous work~\cite{georganas2020ipdps}, we use the $[C_b][N_b][b_n][b_c]$ format for activations since it provides similar performance benefits in the backward by weights training pass.

Our algorithm first assigns the output sub-tensor blocks to the available threads (line 1) and every thread then for each assigned output $Y$ block calculates the addresses of the $W$ and $X$ sub-tensor blocks that need to be multiplied and reduced onto the current output $Y$ block (lines 5-7). Note that our batch-reduce GEMM kernel is JIT-ed, and allows small values of blocking values $b_n$ to be used, and as such we can extract parallelism from the mini-batch dimension even for small values of $N$. By following the loop ordering of Algorithm~\ref{alg:mlp}, a weight sub-tensor is reused by each thread $N_b\_end-N_b\_start-1$ times, potentially from some level of cache. Also, multiple threads are able to read weights from shared caches when the assigned $Y$ blocks correspond to the same subspace of the $K$ dimension. Finally, in case a weight sub-tensor does not fit in the targeted/desired level of cache, we can further block loops at lines 3 and 5. These cache blocking techniques in combination with the flexible blocking factors $b_n$, $b_c$ and $b_k$ which yield high performance micro-kernels, attempt to maximize the data reuse in the GEMM. The backward by data and backward by weights passes follow the same Algorithm~\ref{alg:mlp} since they also instantiate GEMM operations.

\section{Implementation and Optimizations - Multi Socket}
\label{sec:multisocket}
\subsection{Bottom-Up - Multi-socket and multi-node MLP training}
\label{subsec:multisocket_mlp}

A major challenge during multi-socket and multi-node MLP training is to hide the allreduce communication behind the GEMMs of the backward pass. Therefore we study this problem in a
standalone fashion, freed from any framework limitations. 

\begin{figure}[t!]
\centering
\includegraphics[width=0.7\columnwidth]{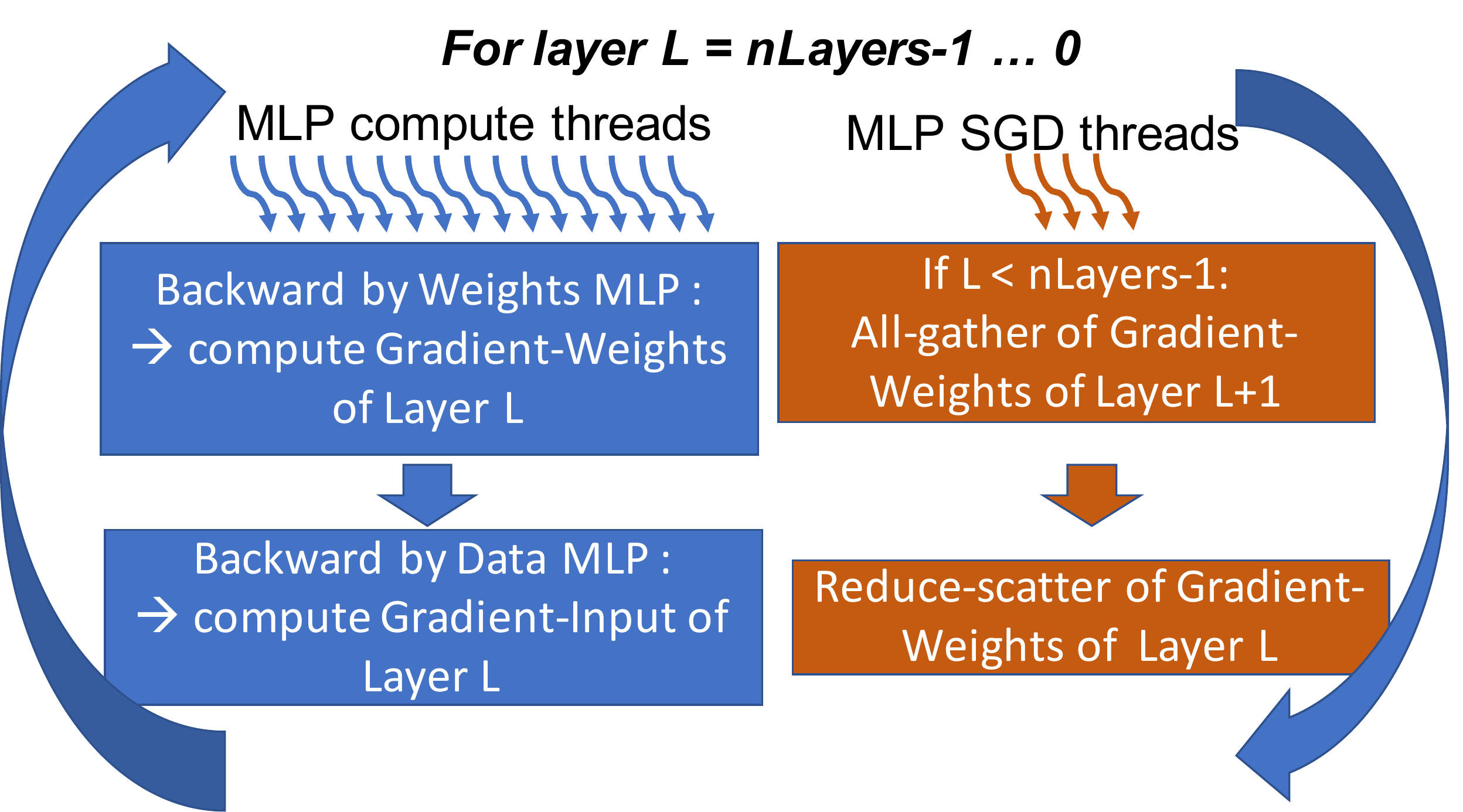}
\caption{Overlapping the SGD solver with the MLP GEMMs in the back-propagation pass of MLP.}
\label{fig:comm_ovlp}
\end{figure}

In order to scale the MLP training to multiple sockets and multiple nodes we adopted the following strategy: First we employ data parallelism (i.e.\ minibatch parallelism) across sockets/nodes, and within each socket we rely on a 2D GEMM decomposition as described in Section~\ref{sec:mlp}. In such a setup, the only communication required across sockets/nodes is during the SGD whereas the 2D decomposition of GEMMs within the socket strives for optimal data movement/reuse.

For the GEMM computations, we leveraged our high-performance Fully Connected/GEMM layers described in subsection~\ref{sec:mlp}. In order to minimize the overhead of the communication in the SGD, we overlapped the SGD solver with the back-propagation MLP kernels. More specifically, the SGD solver can be seen as an All-reduce operation among all the Weight Gradient tensors residing in all sockets/nodes. In our implementation we materialize the all-reduce operation via a reduce-scatter and an all-gather operation. Figure~\ref{fig:comm_ovlp} visualizes how we can overlap the all-gather and the reduce-scatter operations with the compute MLP kernels. Given $T$ cores/threads per socket, we dedicate $S$ threads per socket for the SGD/communication of the gradient weights, and $T-S$ threads for the computation/GEMMs in the back-propagation of the Fully Connected Layers. We tune the value of $S$ in order to balance the communication time in SGD and the computation time in GEMMs. Such MPI progression threads are well-known in large-scale HPC
applications~\cite{7516082,10.1109/IPDPS.2014.113,Sridharan2018OnSD,4663774}.

\subsection{Top-Down - All-to-all Communication in PyTorch Multi-Process}

The original DLRM code from Facebook has support for device-based hybrid parallelization strategies. It follows data-parallelism for MLP and model-parallelism for embeddings. This means that MLP layers are replicated on each agent and the input to first MLP layer is distributed over minibatch dimension whereas embeddings tables are distributed across available agents and each table produces output worth full minibatch. This leads to mismatch in minibatch size at interaction operation and requires communication to align minibatch of embedding output with bottom MLP output. The multi-device (e.g. socket, GPU, accelerator, etc.) implementation of DLRM uses sequence of scatters, one per table, to distribute embedding output over minibatch before entering to interaction operation.  
We extend this parallelization strategy to multi-process using MPI, where a rank can be considered equivalent to a device. We use the Distributed Data Parallel (DDP) module to wrap bottom and top MLPs, whereas for embeddings we simply distribute tables across available ranks. We call this approach ``ScatterList''. One of its drawbacks is that, it makes multiple calls to the communication backend (one call per table). When there are multiple tables per rank this approach turns out to be inefficient, and we optimize it by coalescing output of multiple local tables into one buffer and invoking just one scatter per rank. We call this approach as ``Fused Scatter''. However, if we look at it carefuly, this is a well-known all-to-all communication pattern known to HPC world for years. Nevertheless, DL frameworks such as PyTorch used to lack primitives for supporting this communication pattern. PyTorch has recently added experimental support for alltoall primitive to their distributed backend. We used it in our distributed DLRM code to further minimize the number of calls to communication backend to make it just one call. We call this as ``Alltoall'' version of our code. 

\subsection{Top-Down - Optimal Communication Backend in PyTorch}
To get good scaling efficiency we need to overlap communication with computation. To enable asynchronous communication, the MPI backend of PyTorch spawns a separate thread to drive the communication. The idea is to enable our Bottom-Up approach directly in PyTorch to demonstrate its value in end-to-end runs. The master thread simply enqueues the requested operation to a MPI thread and waits for completion when it is ready to consume the output of communication. While using a separate thread is good for driving communication, as we discussed in previous section, we need multiple threads to saturate the full communication bandwidth. Also, it is very important to manage CPU affinities for this thread. Otherwise it can interfere with compute threads and potentially cause performance degradation. We found that Intel's oneCCL~\cite{ccl2019} library tries to solve these issues. However, it is not integrated into PyTorch as built-in communication backend. PyTorch has recently added an experimental feature that supports adding custom communication backend for torch.distributed module. OneCCL is integrated into PyTorch using this feature~\cite{torch-ccl2020}. We call it as CCL backend and our Alltoall version as ``CCL-Alltoall'' when running with the CCL backend. Finally, for debugging communication performance, we modified DDP module to optionally perform blocking allreduce and added autograd profiling hooks to the communication backends. This allows us to measure time spent in communication primitives which is important for detailed performance analysis.

\section{Hardware Platforms and Benchmarking Setup}
\label{sec:hardware}

As motivated in the introduction we will focus on CPU-based platforms as the 
sparse embedding requires often a huge memory footprint. Due to DLRM's huge memory consumption, 
a single standard dual-socket high-performance computing node is only useful for some
small problems and tuning DLRM's kernels. Therefore we decided to use two different
test beds. The first one is normally used by big database companies such as Oracle or SAP, and is aligned with Facebook's recently announced Zion platform~\cite{Zion2019}: an eight socket 
Skylake Xeon platform. The second
one is a classic HPC configuration: 32 dual-socket Cascade Lake Xeon nodes connected by an Intel 
Omnipath (OPA) pruned fat-tree with two rails per node.

\subsection{8 Socket Shared-Memory Node}

The simplest way of running an application with a large memory footprint is to use 
node with a lot of DRAM. Normally, these nodes are used for databases as the accesses
happen randomly and cannot be easily located. 

\begin{figure}[!t]
  \begin{center}
  \includegraphics[width=0.9\linewidth]{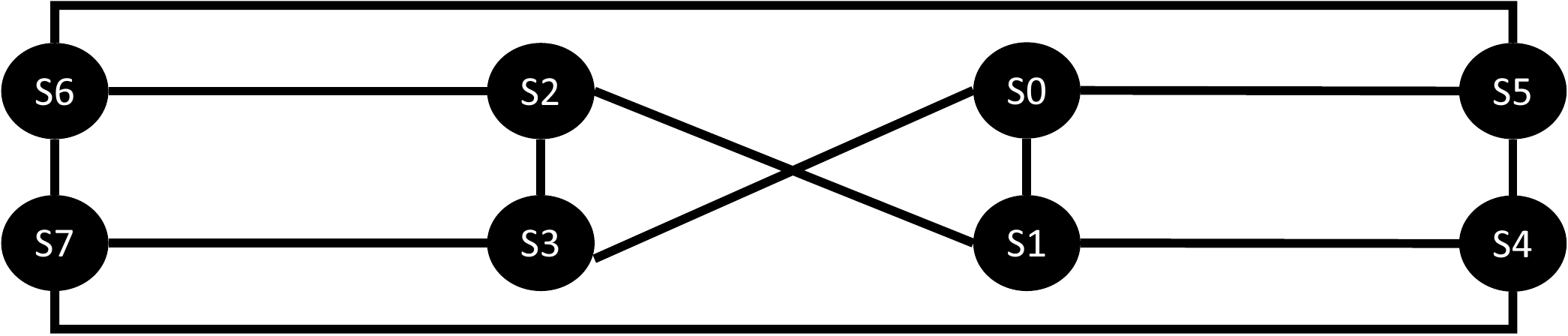}
  \end{center}
  \caption{Schematic of an 8 socket Intel Xeon system, an Inspur TS860M5 machine. As each socket offers only 3 UPI links,
  the sockets are arranged in a twisted hypercube in order to offer a balanced communication
  path between all sockets.}
  \label{fig:8s}
\end{figure}

Figure~\ref{fig:8s} depicts the layout of an Intel Xeon Skylake/Cascade Lake system featuring
8 sockets. The Platinum series processor offers 3 point-to-point Ultra Path Interconnect (UPI) links.
However, each socket needs to communicate with 7 neighbors. Therefore, the sockets are organized
in a twisted hypercube fabric. That ensures that 3 neighbors can be reached in one hop and the remaining 4 
neighbors in two hops. Each of the UPI link offers roughly 22 GB/s bidirectional bandwidth
which is comparable with a
100G fabric in cluster installations, resulting into an aggregated system UPI bandwidth of 
260 GB/s as the machine has 12 unique UPI connections. Each socket is exposed as a NUMA node in the system
such that software can easily optimize the various distances and latencies in case of different local and 
remote NUMA memory accesses. 

In our case, we are using an Inspur TS860M5 machine which has a 2x4 socket node design. Each socket
is equipped with an top-of-the line Intel Xeon Skylake processor (SKX), the Intel Xeon Platinum 8180. 
It features 28 cores at an AVX512 turbo frequency of 2.3 GHz and 1.7 AVX512 base frequency. Therefore 
each socket provides 4.1 TFLOPS FP32 peak performance. With
respect to memory, the system is optimized for high bandwidth and high capacity by using 12 dual-rank 16 GB DDR4-2400
DIMMs per socket offering 100 GB/s memory bandwidth. In total the machine offers 224 core providing 
32\,FP32-TFLOPS at 800 GB/s bandwidth with a capacity of 1.5 TB. 

\subsection{64 Socket Cluster Architecture}

We are going to compare the performance of the 8 socket machine to a traditional HPC installation, namely several dual-node systems connected by low-latency and high-bandwidth fabric, 
in our case Intel OPA.

\begin{figure}[!t]
  \begin{center}
  \includegraphics[width=0.9\linewidth]{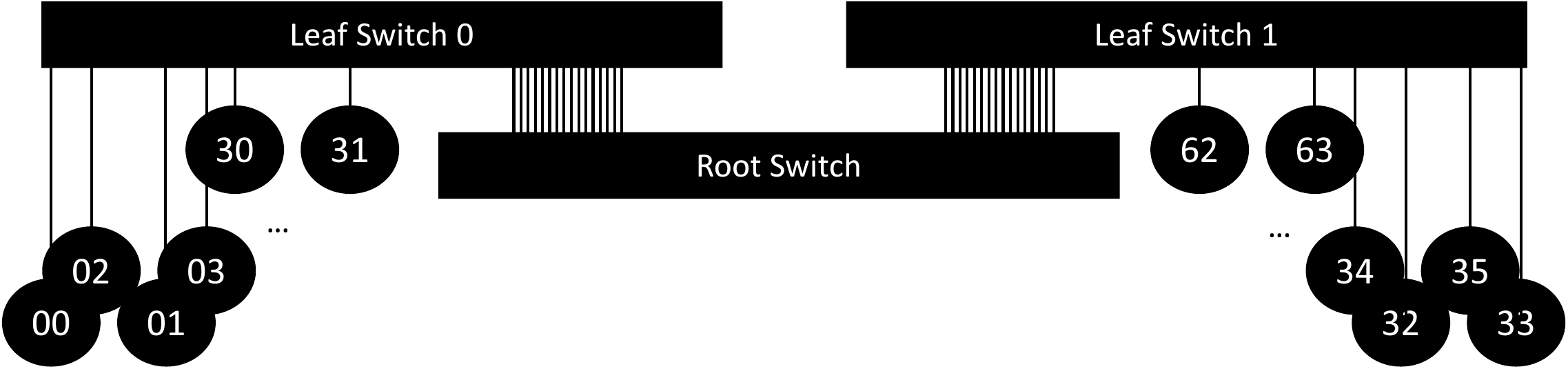}
  \end{center}
  \caption{Schematic of the 64 socket cluster which is connected through a pruned fat-tree. 32 sockets connect with full bandwidth to leaf switches and these two switches are connected with half 
  bandwidth through a root switch.}
  \label{fig:cluster}
\end{figure}

The layout of the cluster we leverage for this experiment, is depicted in Fig.~\ref{fig:cluster}. Each
of the 64 sockets houses its own OPA network adapter which offers 100G connectivity at 1us latency.
As OPA switches allow to connect only 48 endpoints we implemented a classic pruned fat-tree: 16 nodes
with 32 sockets are connected to one switch each, and then both leaf switches are connected with 16 links to
the root switch. Storage and head nodes are also connected to the root switch. This results into 
200 GB/s within each 32 sockets' leaf and 200 GB/s between the leaves as we prune the bandwidth with a 
ratio of 2:1 when going up to the root.

In terms of CPU per socket, the cluster is very similar to the 8 socket system covered earlier. 
Each socket houses a top-bin Intel Cascade Lake processoer (CLX), the Intel Xeon Platinum 8280 which 
offers 100Mhz additional clock over the 8180 of the 8 socket system. 
Its 28 cores run at an AVX512 turbo frequency of 2.4 GHz and 1.8 AVX512 base frequency. Therefore 
each socket provides 4.3 TFLOPS FP32 peak performance. With
respect to memory, the system is optimized for high bandwidth only as its main target is HPC. 
Therefore it only uses 6 dual-rank 16 GB DDR4-2666
DIMMs per socket offering 105 GB/s memory bandwidth. In total the machine offers 1,792 cores providing 
275\,FP32-TFLOPS at 6.7\,TB/s bandwidth with a capacity of 6\,TB. 4 of the 32 nodes have 192GB/socket memory allowing us large single socket runs. 

\subsection{Difference and Similarities Between Both Systems}

As we can see from the two descriptions, we have designed and chosen two very different 
multi-socket systems in terms of technology. However, the application 
visible performance specifications  are
very similar. While OPA offers 100G between sockets, data still needs to be copied through the network
card stack which means multiple internal data copies. In contrast UPI offers as well 100G connectivity
but we can copy data without any additional movements. Therefore on the 8 socket platform, we can
use non-temporal (non-cached) write flows of full cachelines to minimize the communication volume
to a bare minimum. These flows can be regarded as true one-side communication.
\subsection{Benchmarking Setup}
In order to evaluate different pressure points of our hardware systems (covered in the 
following section) we will use three different configurations of DLRM in this work summarized in 
Tab.~\ref{tab:dlrm}. The Small variant is identical to the model problem used in DLRM's
release paper~\cite{Naumov2019DeepLR}. Large variant is the small problem scaled in every aspect for 
scale-out runs. The MLPerf configuration is recently proposed as a benchmark config for performance evaluation of a recommendation system training~\cite{mlperf}. 
It uses Criteo Terabyte Dataset~\cite{terabyte} for its convergence run. For single socket run, we use \texttt{numactl} to bind it to socket 0 whereas distributed run always use 1 socket per rank and occupy the node first before going multiple nodes.
\begin{table}[!t]
\centering
\begin{tabular}{c||c|c|c}
  \textbf{Configuration Parameter} & \textbf{Small}  &  \textbf{Large} & \textbf{MLPerf}\\
  \hline
  Minibatch (MB) (Single Socket) ($N$) & 2048 & - & 2048\\
  Global MB for Strong Scaling ($GN$) & 8192 & 16384 & 16384\\
  Local MB for Weak Scaling ($LN$) & 1024 & 512 & 2048\\
  Avg. look-ups per Table ($P$) & 50 & 100 & 1\\
  Number of Tables ($S$) & 8 & 64 & 26\\
  Embedding Dimension ($E$) & 64 & 256 & 128\\
  \#rows per table ($M$) & $1 \cdot 10^6$ & $6 \cdot 10^6$ & Up to 40M\\
  Length Inputs Bottom MLP & 512 & 2048 & 13\\
  \#Layers Bottom MLP & 2 & 8 & 3\\
  Bottom MLP Size & 512 & 2048 & 512-256-128\\
  \#Layers Top MLP & 4 & 16 & 4\\
  Bottom MLP Size & 1024 & 4096 & 512-512-256-1\\
\end{tabular}
\vspace{2ex}
\caption{DLRM model specifications used in this work. Small is taken from the release paper of 
DLRM~\cite{Naumov2019DeepLR}.}
\label{tab:dlrm}
\end{table}

\section{Analysis and Results}
\label{sec:results}

In this Section we present the performance results for the aforementioned 
implementations. We start with a bottom-up analysis and end with 
large-scale cluster experiments. Each result is the average of multiple batches of iterations to account
for potential small run-to-run performance variations.

\subsection{Bottom-Up: Single Socket standalone MLP results}
\begin{figure}[t!]
\centering
\includegraphics[width=\columnwidth]{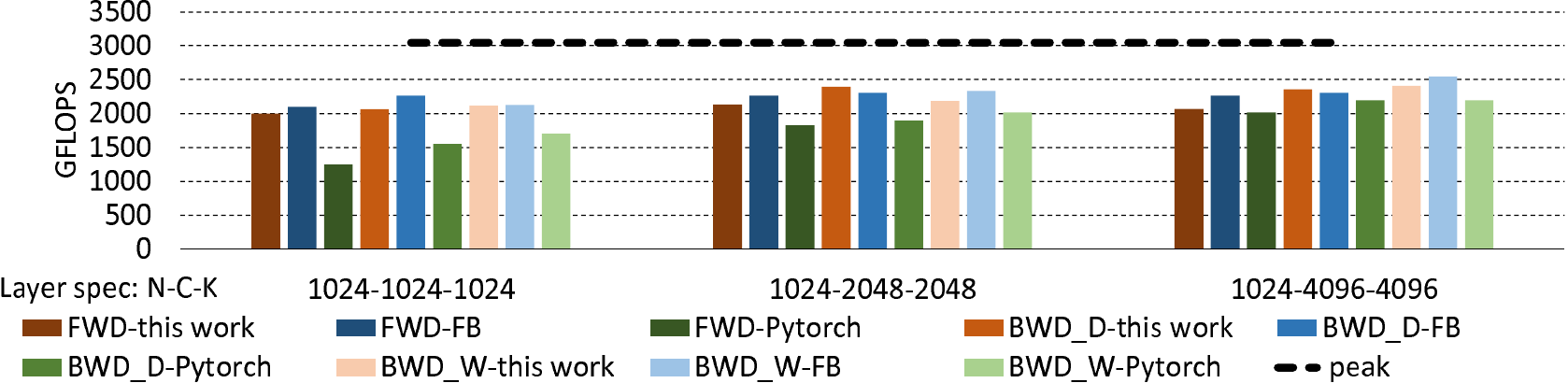}
\caption{MLP training kernel performance on single SKX socket}
\label{fig:mlp_singlesocket}
\end{figure}
Figure~\ref{fig:mlp_singlesocket} shows the performance of 5-layer MLP training kernels on a single SKX socket. In particular we compare all three training passes, namely forward (FWD), backward by data (BWD\_D) and backward by weights (BWD\_W) for three different implementations: i) this work (orange-shaded bars), ii) an MLP training code optimized by Facebook~\cite{fbmlp} for the SKX platform (blue-shaded bars) and iii) the PyTorch implementation that leverages multi-threaded MKL calls for the involved GEMMs (green-shaded bars). Also, we experimented with various MLP configurations, where we fixed the minibatch $N$ to be 1024 and we varied the feature map dimensions $C$ and $K$ (we tried the values 1024, 2048 and 4096). The Facebook MLP implementation is similar to ours: it employs a NUMA-aware and thread-aware blocking of the Fully Connected layers/GEMMs and for the single-threaded kernels it uses serial MKL GEMM calls. We observe that our approach and Facebook's implementation show similar performance on a single socket: the average performance across all configurations and all passes is $72\%$ and $75\%$ of peak respectively. On the contrary, the MLP implementation in PyTorch which employs large, multi-threaded MKL calls shows average efficiency 61$\%$ of peak and is $\sim 18\%$ slower than our approach.

\subsection{Bottom-Up: Multi Socket / Multi-node standalone MLP results}
\label{subsec:distMLP}

 \begin{figure}[t!]
 \centering
 \includegraphics[width=0.75\columnwidth]{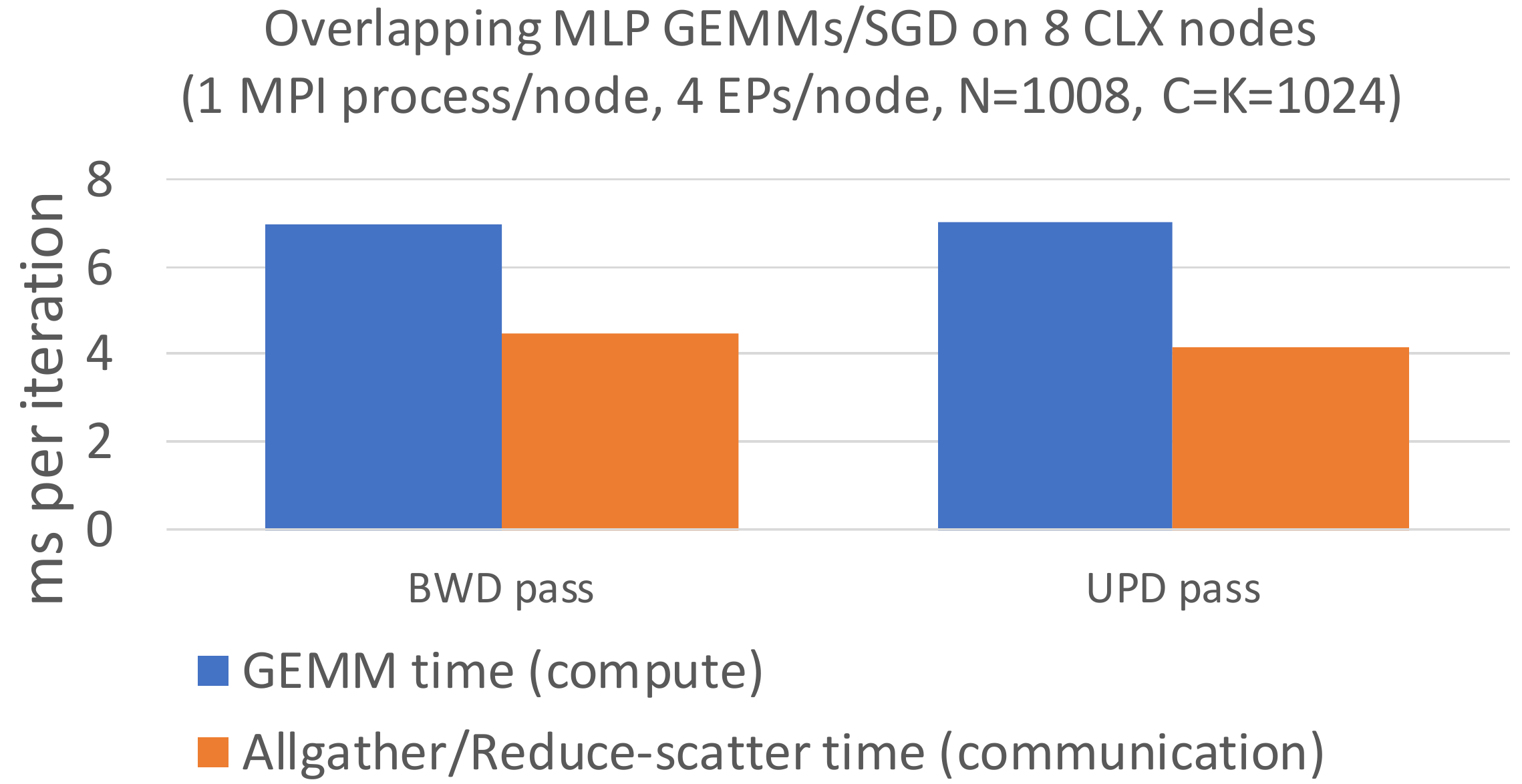}
 \caption{Communication/Computation overlap on 8 CLX nodes (1 MPI process/node, N=1008, C=K=1024)}
 \label{fig:comm_ovlp_multinode}
 \end{figure}

We evaluate the efficacy of the computation/communication overlapping during the standalone MLP training implementation. For the 8 socket shared memory node, we implemented the SGD via non-temporal flows, and the communication takes place over the UPI connections between the sockets. More specifically, we dedicated 4 cores per socket for the SGD communication flow which is overlapped with the back-propagation GEMMs that leverage 24 cores per socket. Similarly, for the multinode setup, we assign one MPI-rank per socket, dedicated 24 cores per socket for the back-propagation GEMMs whereas we created 4 MPI Endpoints (EPs) per socket for the communication. As we see on Figure~\ref{fig:comm_ovlp_multinode}, such a setup/problem configuration is sufficient to completely hide the communication (orange bars) behind the computations/GEMMs (blue bars). It is also worth-noting that our overlapping scheme can hide the SGD/communication cost over UPI (i.e.\ in the 8 socket node). E.g.\ for the smallest config of Figure~\ref{fig:mlp_singlesocket} the backward by data and backward by weights GEMMs take 5.40 and 5.39 ms respectively while the corresponding communication operations that are overlapped require 2.84 and 1.86 ms.

\subsection{End-to-End: Single Socket DLRM results}
\begin{figure}[!t]
  \begin{center}
  \includegraphics[width=0.9\linewidth]{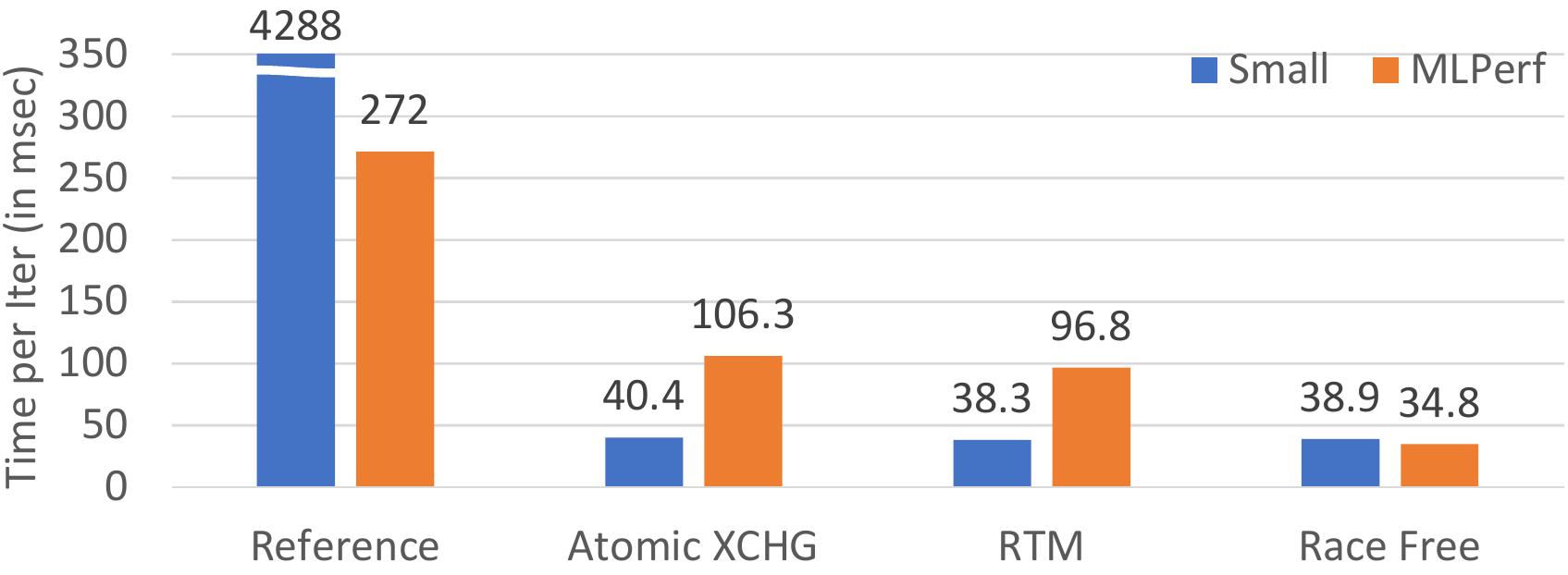}
  \end{center}
  \caption{DLRM single socket performance}
  \label{fig:single_node}
\end{figure}

\begin{figure}[!t]
  \begin{center}
  \includegraphics[width=0.9\linewidth]{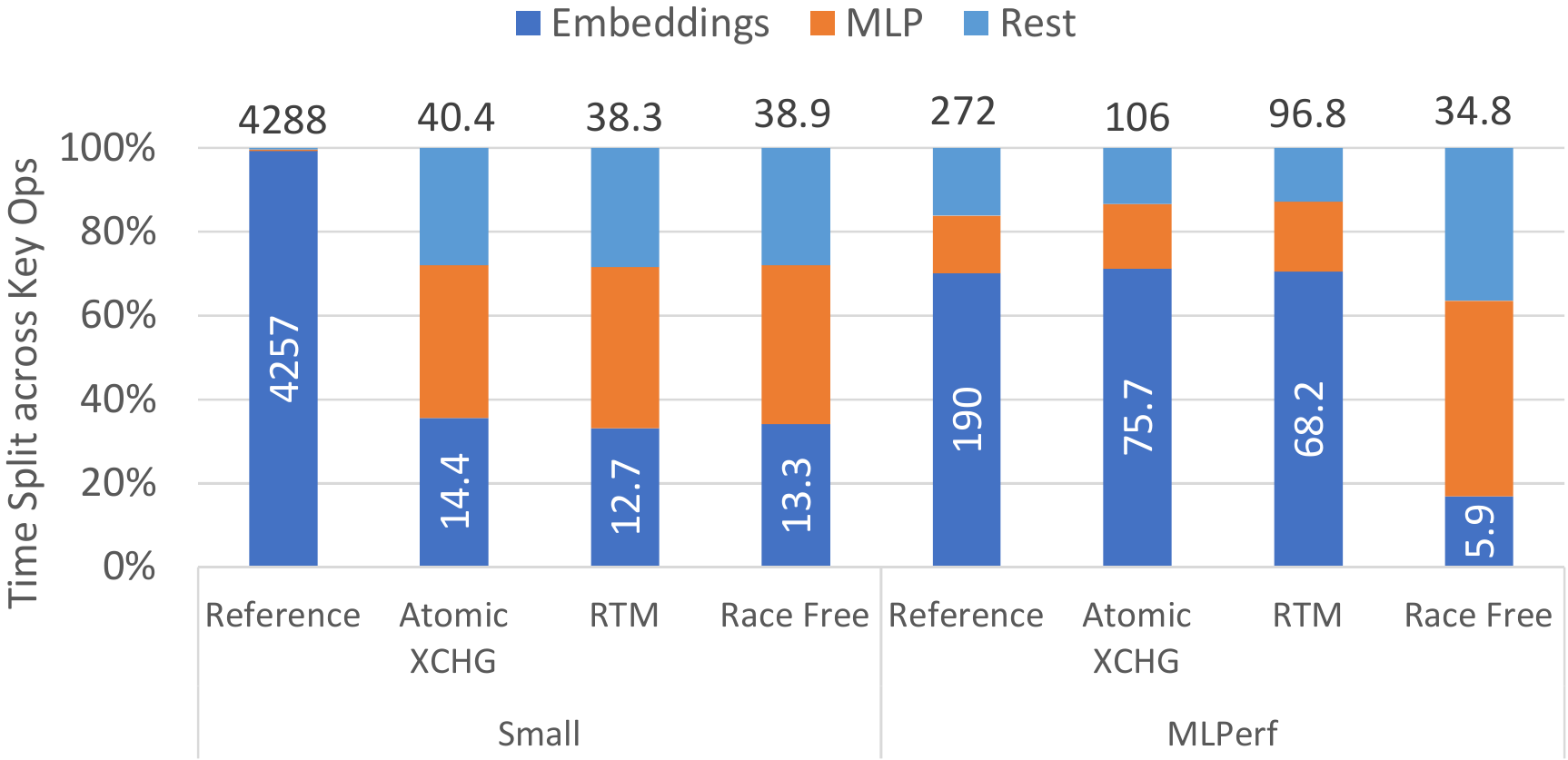}
  \end{center}
  \caption{DLRM single socket performance breakdown}
  \label{fig:single_node_breakdown}
\end{figure}

We started our DLRM analysis using the latest release of PyTorch v1.4.0. After analyzing profiling data for each operation of the DLRM code, we found that 99\% of the execution time was spent in just one kernel related to the sparse embedding look up.
This was caused by a naive CPU backend implementation which was focused on functionality instead of performance. We found few more similar instances but performance impact was dominated by a single embedding update kernel. 
Therefore, we carefully reviewed all PyTorch framework operations involved when executing DLRM. While trivial issues were fixed in PyTorch source code, we implemented a PyTorch c++ extension for EmbeddingBag operations for quick experimentation and implemented various update strategies described in Sec.~\ref{sec:singlesocket}. The performance benefits depend on config and ranges from 110$\times$ for small config to 8$\times$ for the MLPerf config. We could not run the large config on single socket as it needs minimum of 450GB DRAM memory capacity. Figure \ref{fig:single_node_breakdown} shows percent of time spent in various key components before and after applying optimizations. We can see that for the small config, all the 3 embedding update strategies perform more or less same as indices distribution is random and have very little contention. However, for the MLPerf config, we observe a lot of contention with the terabyte dataset causing up to 10$\times$ slowdown for embedding update operation compared to our race free algorithm. We also observe that even though it appears (with a first glance) that DLRM performance is embedding-heavy, after optimizations, it takes about 30\% of total time for small config matching it with MLP time whereas for the MLPerf config, embeddings take less than 20\% of total time.

Last but not least, we can summarize that the small config, runtime per iteration decreased from 4288ms to 38ms. Additionally, we want to mention that previous work~\cite{Naumov2019DeepLR} timed an NVIDIA V100 GPU at 62ms running  exactly the same DLRM problem using the Caffe2 frontend. However, to be fair, we
have to mention that V100 has roughly 3.5$\times$ more FP32-FLOPS than Skylake/Cascade and 8$\times$ more available bandwidth at much smaller memory capacity. Therefore, we can expect a fully-optimized 
GPU software stack to be at around 10-15ms for the small problem, being 2-3x faster than our optimized single-socket CPU version. Nevertheless, that puts the CPU into a very strong position as it can run virtually any problem configuration without memory limitations as we will show in the next section. 
More importantly, a CPU-cluster is not limited by small memory capacity per socket and therefore the sweet spot between MLP and embedding performance can be easily picked.

\subsection{End-to-End: Multi Socket / Multi-node DLRM results}
As a part of distributed DLRM scaling analysis, we wanted to find answers to following  questions and reason about the performance we see.
\begin{enumerate}
    \item What is the pure compute scaling?
    \item What is the cost of pure communication when there is no compute interference?
    \item How does the allreduce and alltoall communication time change as we increase the number of ranks?
    \item What is the impact of overlapping communication?
    \item How much communication time can be overlapped?
    \item How much time is spent in preparing  actual communication
    such as copy to flat buffers or gradient averaging?
\end{enumerate}

\begin{table}[!t]
\centering
\begin{tabular}{c||c|c|c}
  \textbf{Parameter} & \textbf{Small}  &  \textbf{Large} & \textbf{MLPerf}\\
  \hline
  Mem capacity required for all tables (in GB) & 2 & 384 & 98\\
  Minimum sockets required  & 1 & 4 & 1*\\
  Maximum Ranks to scale & 8 & 64 & 26\\
  Total AllReduce Size (in MB) & 9.5 & 1047 & 9.0\\
  Strong Scaling Alltoall Volume (in MB) & 15.8 & 1024 & 208\\
\end{tabular}
\vspace{2ex}
\caption{DLRM model characteristics for distributed run}
\label{tab:dlrm_comms}
\end{table}

We evaluated the multi-node/multi-socket performance of DLRM for all the 3 configs on our 64 socket cluster and an 8 socket shared-memory node. We performed strong scaling experiments to size reduction in time-to-train as we leverage more sockets to solve a fixed problem. We also performed weak scaling analysis to understand the pure communication overheads as we scale on multiple sockets. Due to pure model parallelism used for distributing embedding tables, the maximum number of ranks we can use to scale a config depends on the total number of embedding tables we have in the config. %Table~\ref{tab:dlrm_comms}.
Since, the small config has only 8 embedding tables, we can scale it on up to 8 sockets. Large config can be scaled on up to 64 sockets while MLPerf can be scaled on up to 26 sockets. Moreover, since the large config has a large memory foot print, we can only start running it with minimum of 4 sockets. Therefore, we use 4 ranks best performance (CCL-Alltoall version) as our baseline for computing efficiency or speed up for the large config. For the other two configs, we use optimized single socket performance as a baseline for comparison.
Before looking at scaling performance we want to understand the communication characteristics of the 3 configs and set some expectations.
\begin{equation}
SZ_{\text{allreduce}} = \sum\limits_{l=0}^n f_{i}^l\times f_{o}^l + f_{o}^l
\label{eq:allreduce}
\end{equation}
\begin{equation}
SZ_{\text{alltoall}} = S \times N \times E
\label{eq:alltoall}
\end{equation}
Eq.~\ref{eq:allreduce} shows the size for allreduce as seen by each rank where $f_{i}^l$ and $f_{o}^l$ are the input and output feature maps for given MLP layer and $n$ is total number of layers in top and bottom MLP. $SZ_{\text{allreduce}}$ is independent of number of ranks or minibatch size. Therefore, cost of allreduce increases steadily as we increase the number of ranks and imposes a major challenge for strong scaling compared to weak scaling. Eq.~\ref{eq:alltoall} shows total volume of alltoall communication, now across all the ranks, related to embedding tables communication. This volume is proportional to the global minibatch ($N$) and thus remains constant for strong scaling and increases proportional to number of ranks for weak scaling. Since, this volume gets divided across all the ranks, the size of point-to-point message reduces $4\times$ as we double the number of ranks. Also, alltoall directly benefits multiple links such as UPI across sockets and OPA across nodes. Similarly, cost of alltoall is expected to reduce by 4x when going from 2 to 4 ranks and then steadily go down as we increase number of ranks further.
Table~\ref{tab:dlrm_comms} shows the allreduce buffer size for MLPs and alltoall volume for the strong scaling configs we used in our evaluation. Looking at these parameters, we expect the small and large problem to be allreduce-bound whereas the MLPerf config would initially be alltoall-bound and becomes allreduce-bound for high rank counts. Finally, since the output of allreduce is used at the end of backward pass, it can be overlapped over entire compute of backward pass (see Section~\ref{subsec:multisocket_mlp}) whereas alltoall can only be overlapped with bottom MLP compute. Thus, cost of alltoall is more difficult to hide compared to allreduce cost. 
In order to answer the questions 1 to 6 effectively, we instrumented the PyTorch source code to optionally perform blocking communication for allreduce and alltoall (shown in the ``blocking'' part of the graphs in Figure \ref{fig:cc_brk_strong}, \ref{fig:comm_brk_strong}, \ref{fig:cc_brk_weak} \& \ref{fig:comm_brk_weak}). The bars in ``overlapping'' part shows compute and exposed communication time as seen by application after overlap. 
\subsubsection{\textbf{Strong Scaling}}
\begin{figure*}[!t]
  \begin{center}
  \includegraphics[width=0.45\linewidth]{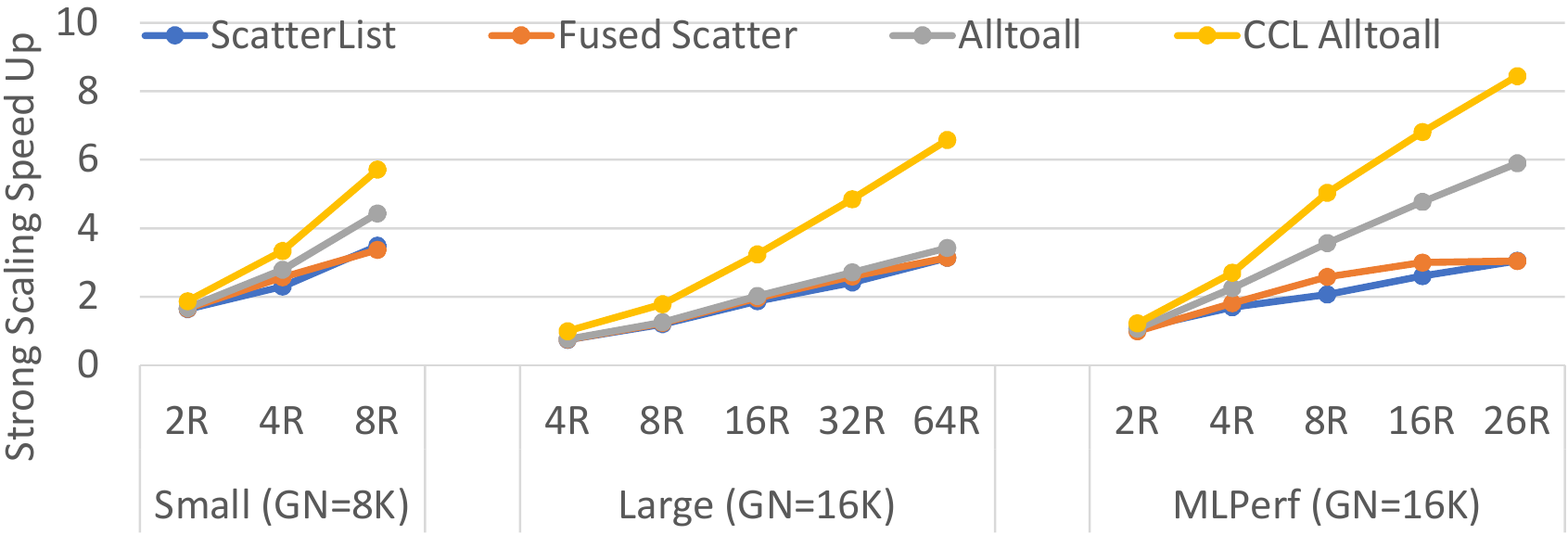}
  \hspace{0.05\linewidth}
  \includegraphics[width=0.45\linewidth]{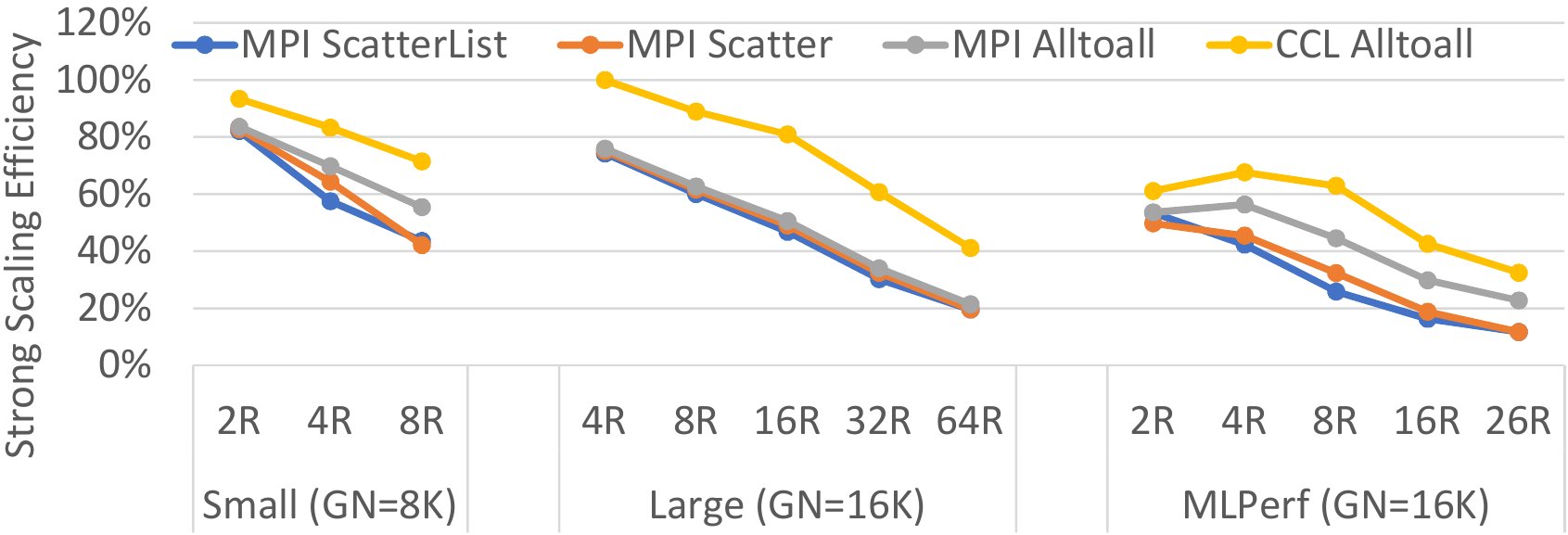}
  \end{center}
  \caption{DLRM strong scaling performance wrt. optimized baseline (Left: speed-up, Right: scaling efficiency)}
  \label{fig:strong_scaling_clx}
\end{figure*}

\begin{figure*}[!t]
  \begin{center}
  \includegraphics[width=0.45\linewidth]{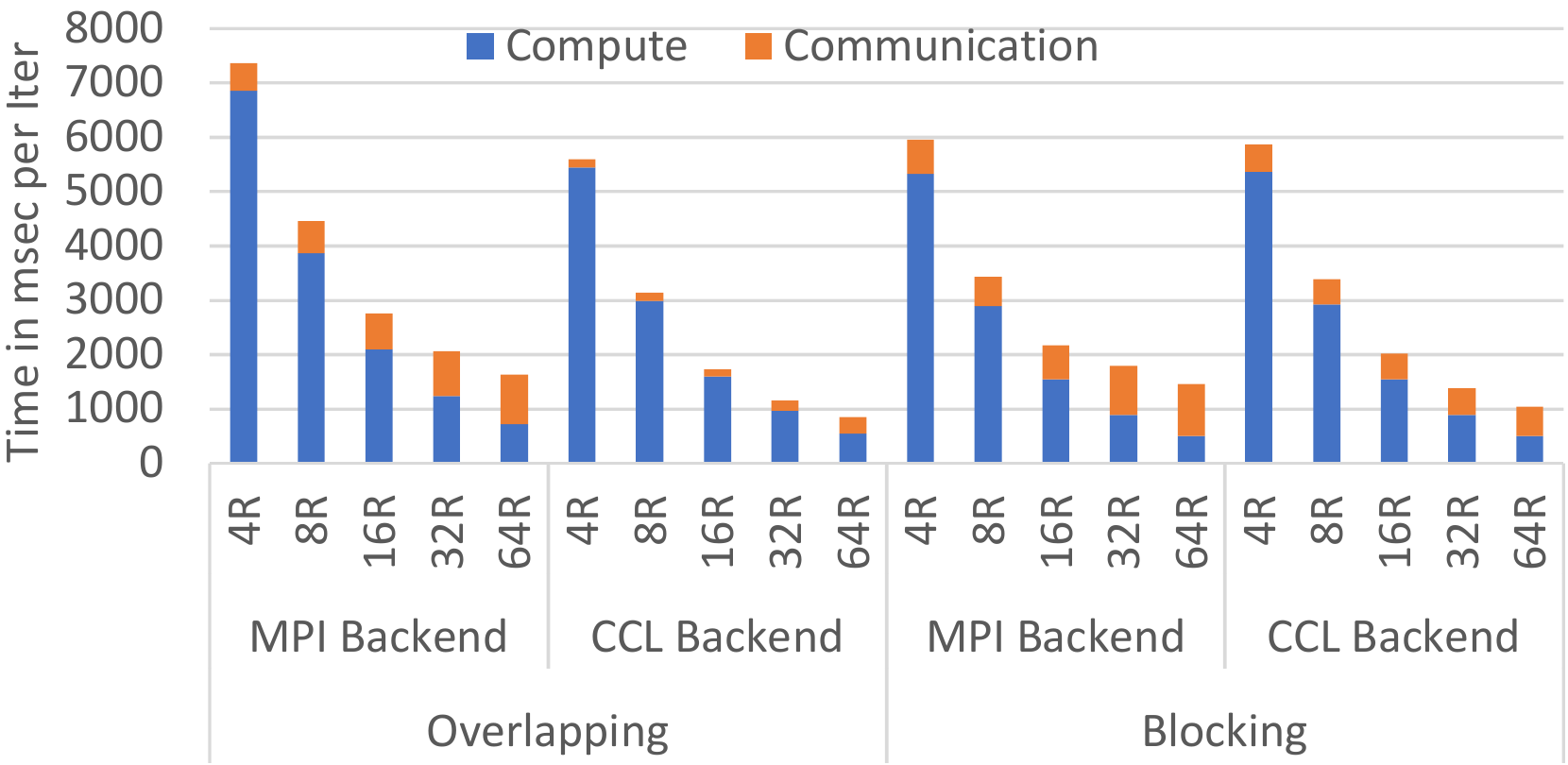} \hspace{0.05\linewidth}
  \includegraphics[width=0.45\linewidth]{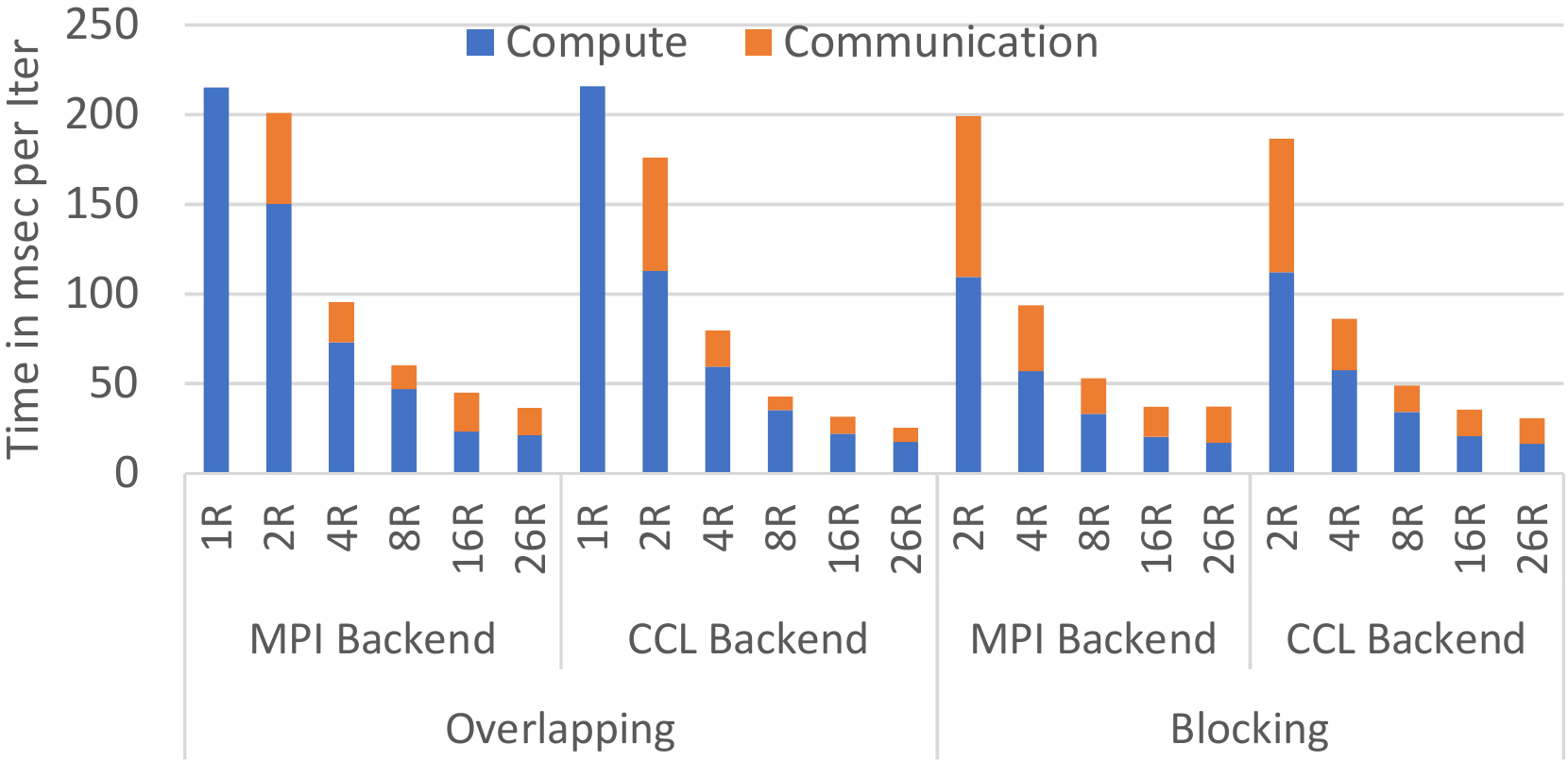}
  \end{center}
  \caption{Compute-Communication time break up with and without overlap for strong scaling (Left: Large Config, Right: MLPerf Config)}
  \label{fig:cc_brk_strong}
\end{figure*}

\begin{figure*}[!t]
  \begin{center}
  \includegraphics[width=0.45\linewidth]{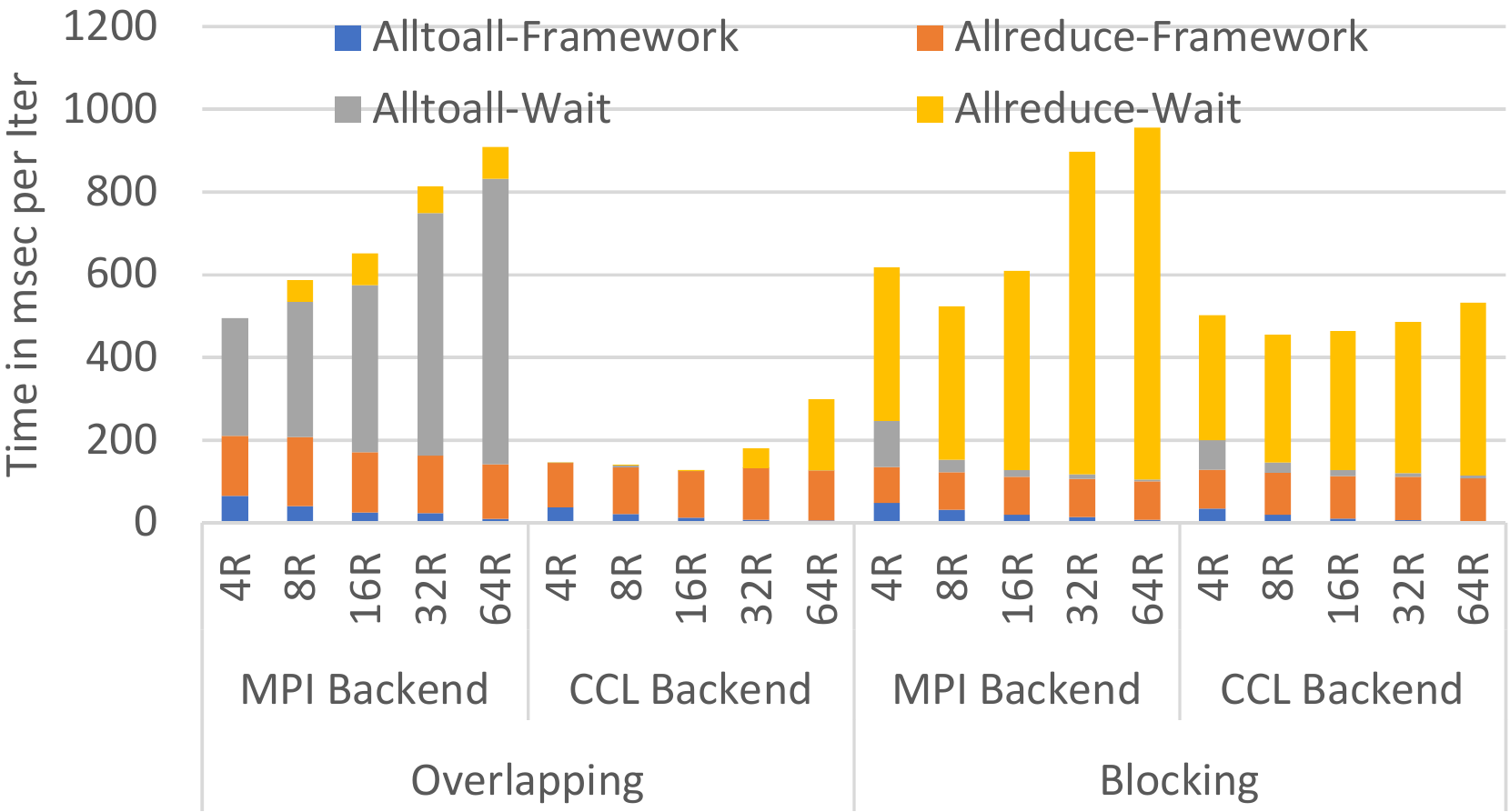}
  \hspace{0.05\linewidth}
  \includegraphics[width=0.45\linewidth]{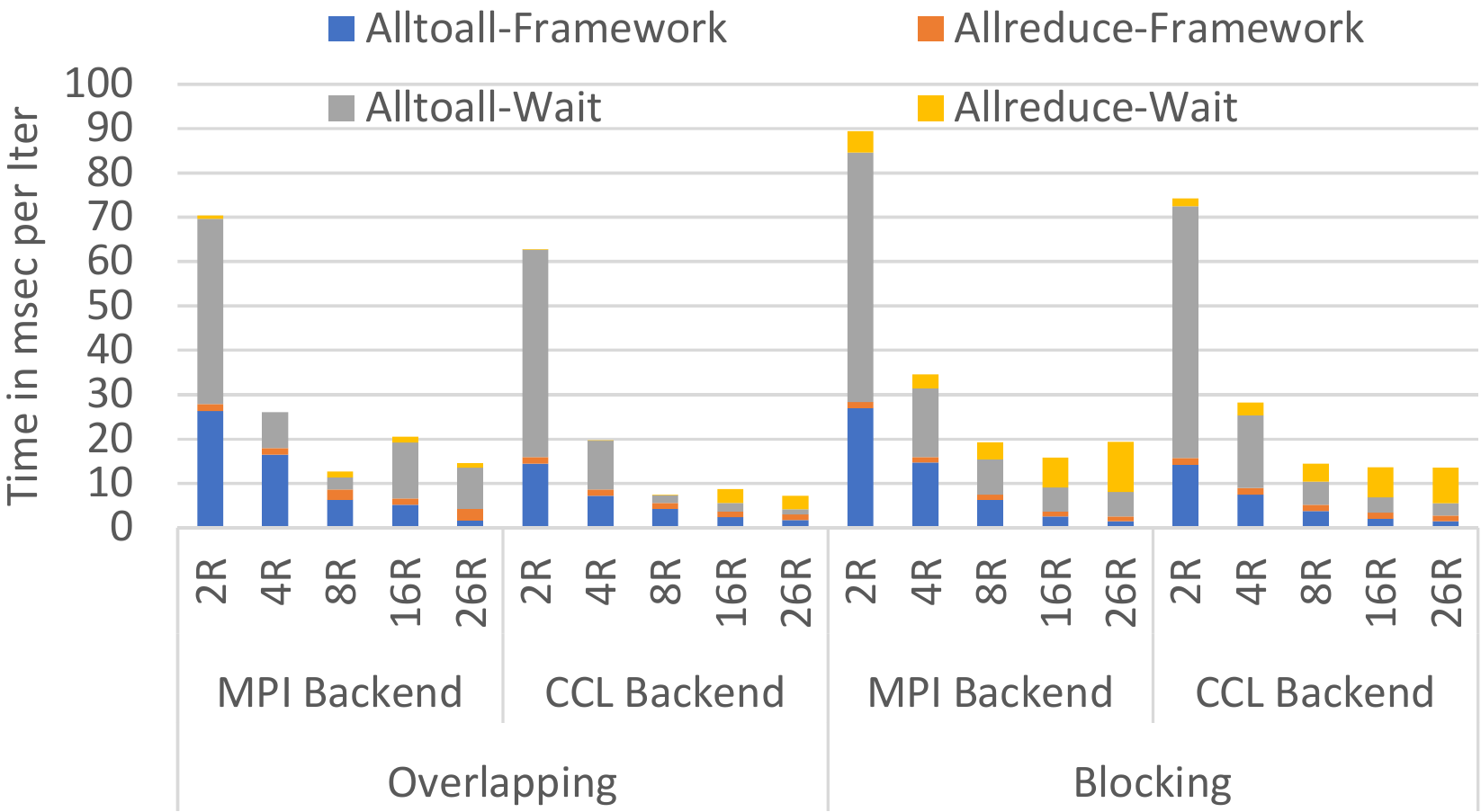}
    \end{center}
  \caption{Communication time break up with and without overlap for strong scaling (Left: Large Config, Right: MLPerf Config)}
  \label{fig:comm_brk_strong}
\end{figure*}

Figure~\ref{fig:strong_scaling_clx} 
shows strong-scaling speed-up and efficiency of scaling when scaled up to 64 ranks (64R). We achieve up to 8.5x end-to-end speed up for the MLPerf config when running on 26 sockets (33\% efficiency) and about 5x-6x speed up when increasing the number of sockets by 8x for the small and large configs ($\sim$60\%-71\% efficiency). It is obvious that the native alltoall performs better than scatter-based alltoall implementations and yields more than 2x performance benefits. However, even MPI-based alltoall leaves a lot of performance on the table as it becomes evident from performance of CCL-Alltoall version which gives up to 1.4x additional speed up for end-to-end time. To understand this in more detail, we look at the compute-communication breakdown for large and MLPerf configs as shown in 
Figure~\ref{fig:cc_brk_strong}.
Here we skip the analysis of small config which has a similar behavior to that of the large problem. As we can see, when using the MPI-backend, not only communication time is higher than for the CCL-backend but even compute times grow significantly. This is not intuitive. After looking into individual kernels' performance, we observed that almost all compute kernels were slowed down due to communication overlap. This was happening because the thread spawned by PyTorch MPI backend for driving MPI communication was interfering with compute threads and slowing down both the compute and communication. To avoid this, CCL provides a mechanism to bind the communication threads to specific cores and we exclude these cores from compute cores. Therefore, we do not see compute slowdown issues but much better overlap of compute and communication with the CCL-backend. 
Figure~\ref{fig:comm_brk_strong} depicts further breakdown of communication cost as time spent in pre- and post-processing for actual communication and actual time spent in wait calls. The pre- and post-processing costs remain comparable across both backends. However, even the pure cost of communication is lower with CCL-backend compared to the MPI-backend. This is because CCL uses multiple cores to drive the communication. 
Another puzzling thing we observed for large problem is the cost of alltoall. We see a huge alltoall cost for MPI backend when overlapping communication is used but almost negligible cost when blocking communications are used. After careful analysis, we concluded that this is happening due to in-order completion nature of MPI-backend that shows up as cost of allreduce at alltoall wait.    

Given all this, we see good initial efficiency of scaling for small and large configs which drops steadily due to exposed allreduce cost as we increase the number of ranks. For the MLPerf config, the profile starts with lower efficiency at 2 ranks due to very high alltoall cost. Then initially, efficiency goes up until 8 ranks as cost of alltoall reduces and again drops at 16 and 26 ranks as allreduce cost starts to assemble the majority of communication. This is in line with our expectations for our simple performance with Eq.~\ref{eq:allreduce} and Eq.~\ref{eq:alltoall}. 

\subsubsection{\textbf{Weak Scaling}}
\begin{figure*}[!t]
  \begin{center}
  \includegraphics[width=0.45\linewidth]{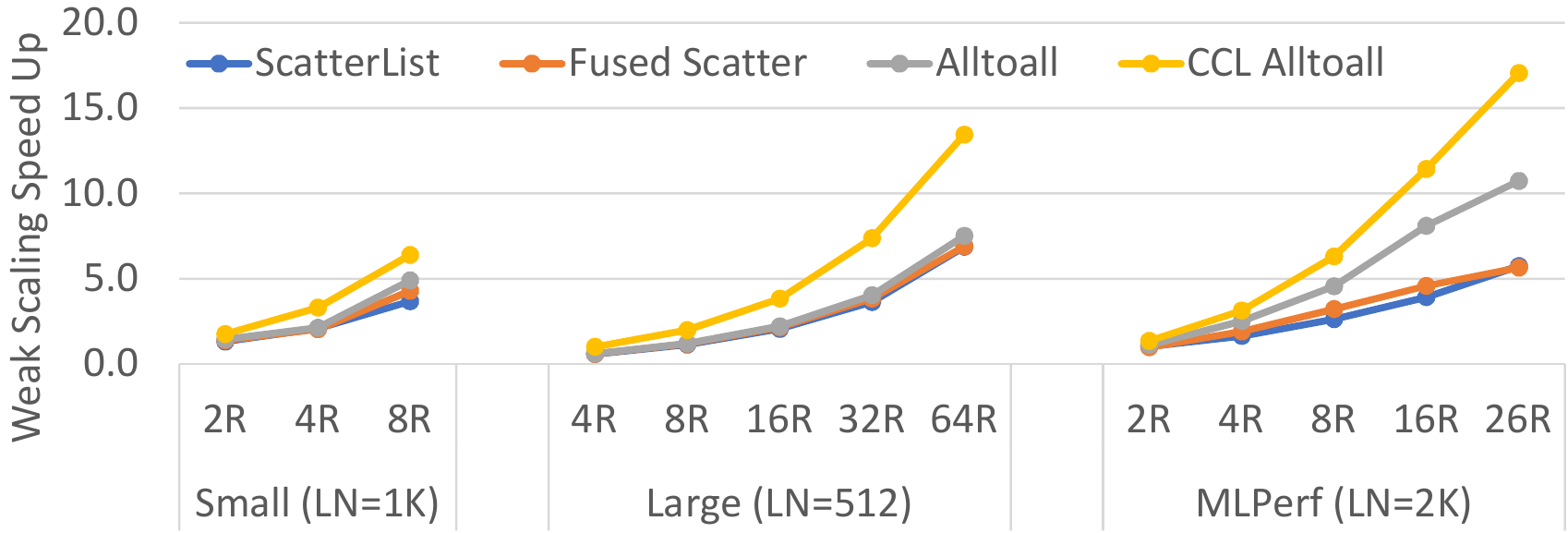}
  \hspace{0.05\linewidth}
  \includegraphics[width=0.45\linewidth]{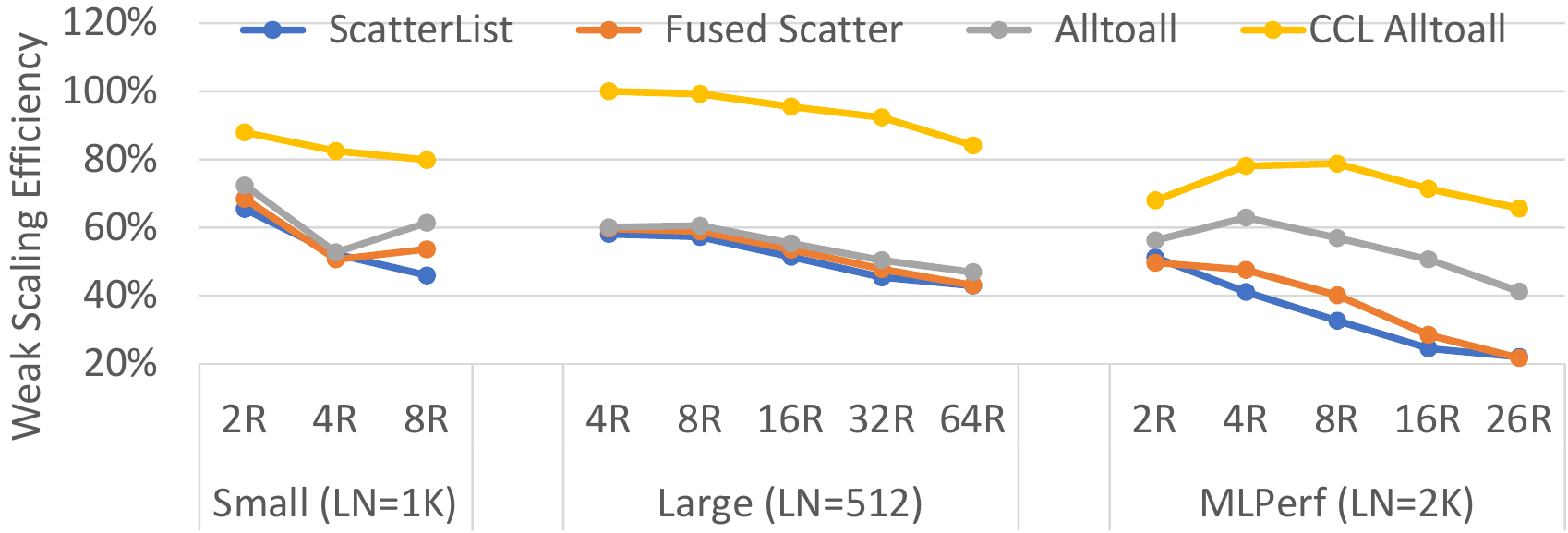}
  \end{center}
  \caption{DLRM weak scaling performance wrt. optimized baseline (Left: speed-up, Right: scaling efficiency)}
  \label{fig:weak_scaling_clx}
\end{figure*}

\begin{figure*}[!t]
  \begin{center}
  \includegraphics[width=0.45\linewidth]{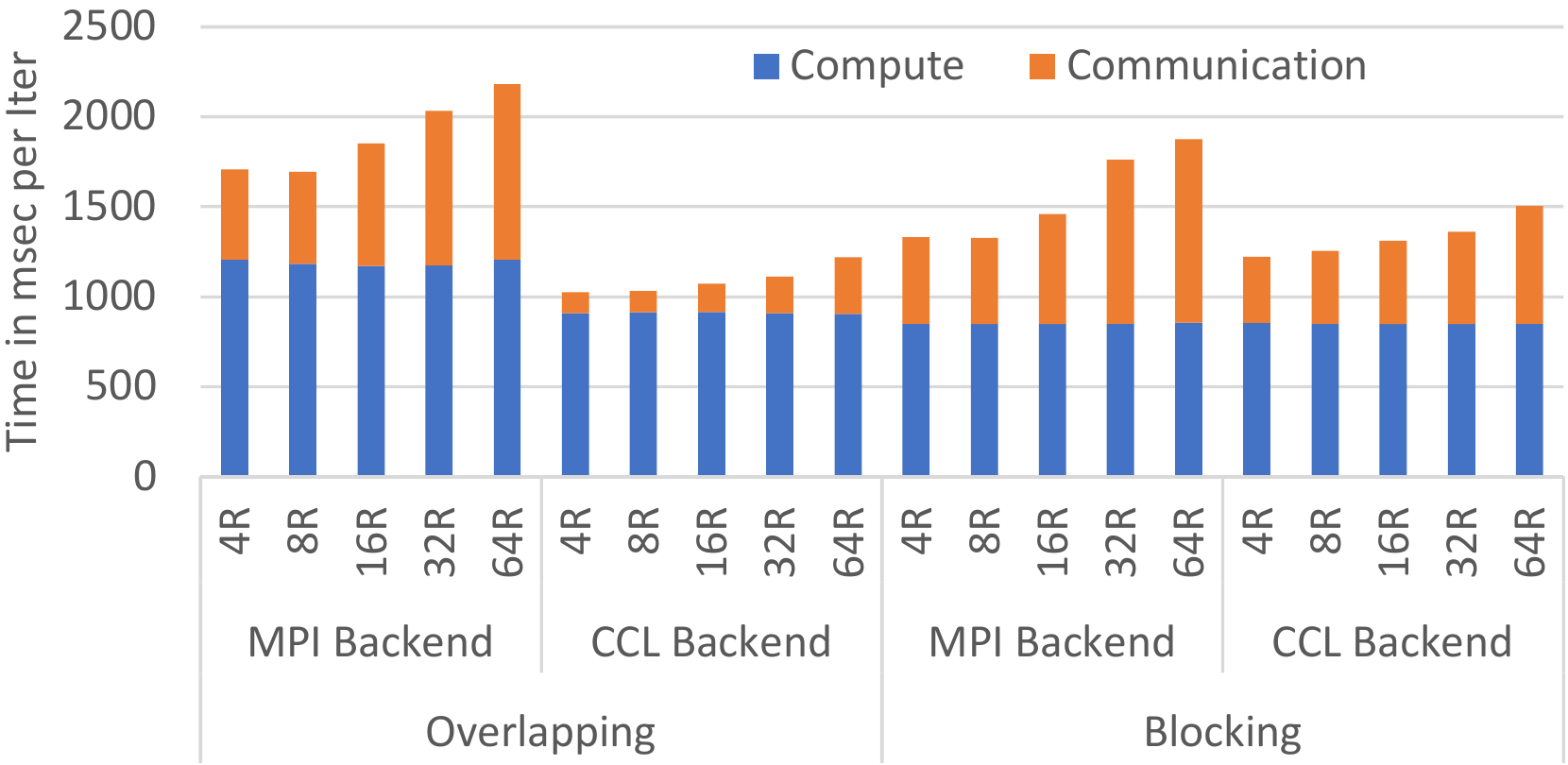} \hspace{0.05\linewidth}
  \includegraphics[width=0.45\linewidth]{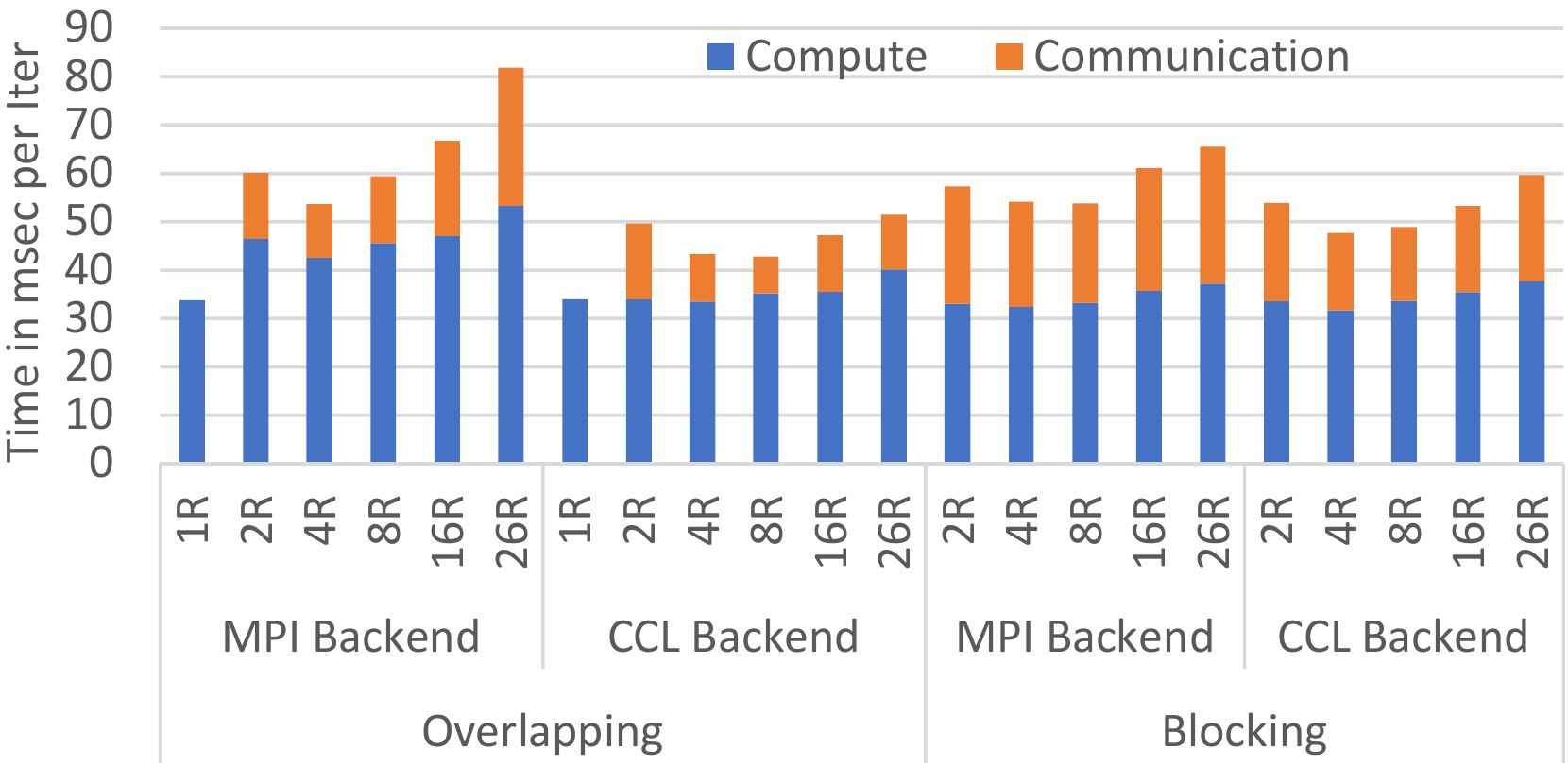}
  \end{center}
  \caption{Compute-Communication time break up with and without overlap for weak scaling (Left: Large Config, Right: MLPerf Config)}
  \label{fig:cc_brk_weak}
\end{figure*}

\begin{figure*}[!t]
  \begin{center}
  \includegraphics[width=0.45\linewidth]{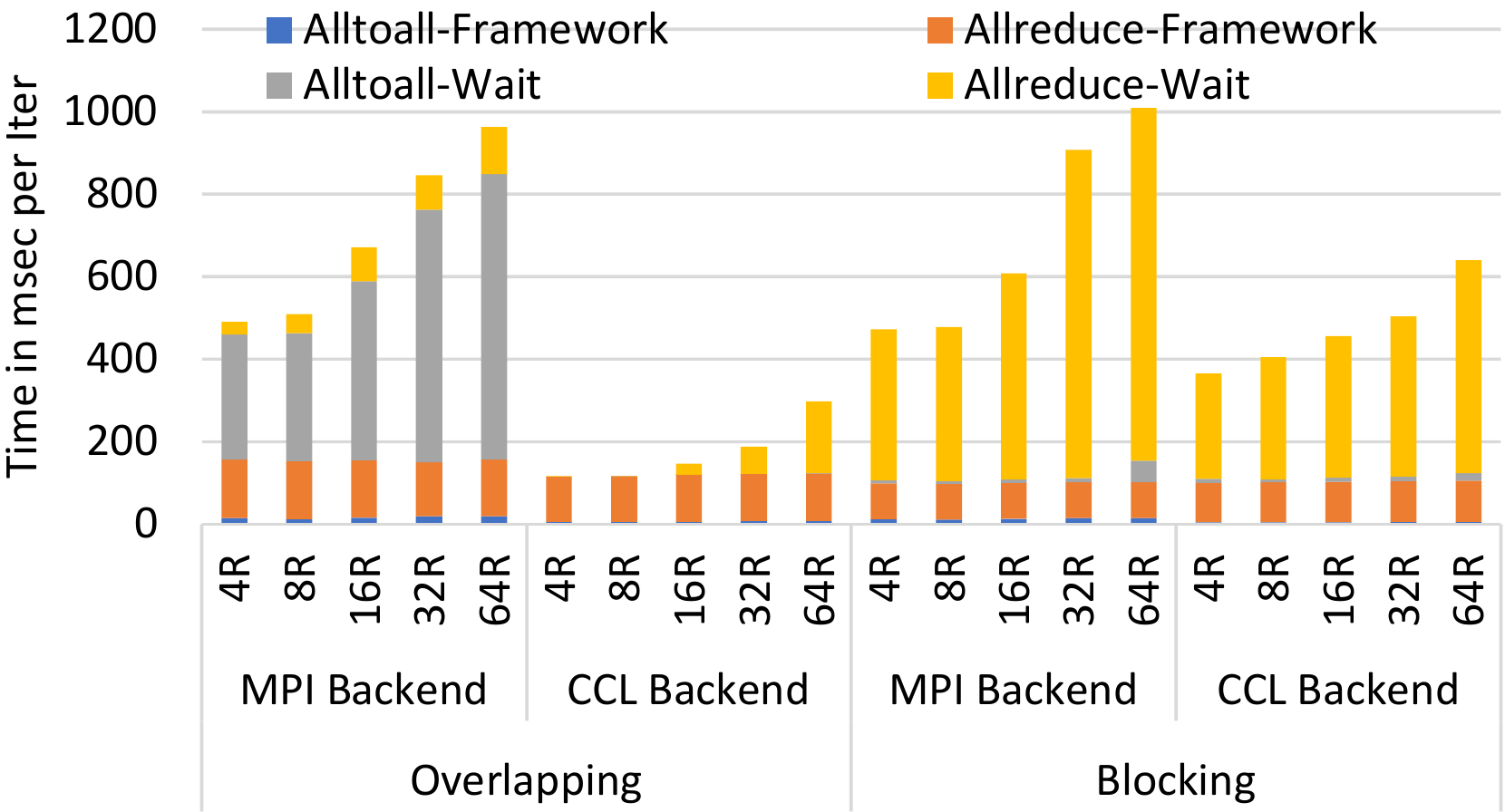}
  \hspace{0.05\linewidth}
  \includegraphics[width=0.45\linewidth]{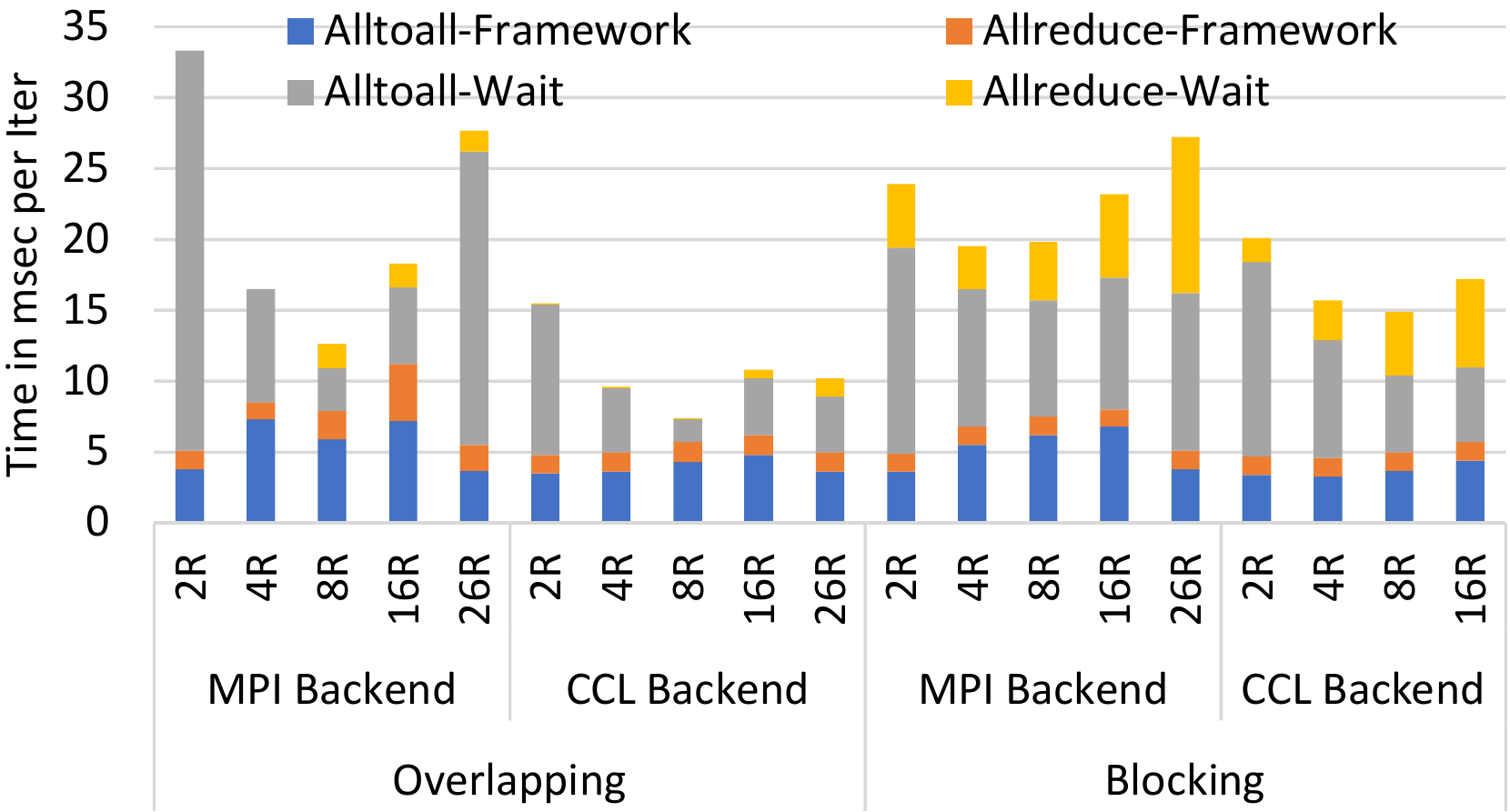}
    \end{center}
  \caption{Communication time break up with and without overlap for weak scaling (Left: Large Config, Right: MLPerf Config)}
  \label{fig:comm_brk_weak}
\end{figure*}

Figure~\ref{fig:weak_scaling_clx} 
shows weak-scaling speed-up and efficiency. With weak-scaling, we achieve up to 17x end-to-end speed up for the MLPerf config when running on 26 sockets (65\% efficiency) and 13.5x speed up (84\% efficiency) when increasing number of sockets by 16x (64 ranks compared to 4 ranks optimized baseline) for the large config. In case of the small config, we achieve about 6.4x speed up on 8 sockets (80\% efficiency). We see a similar trend for using native alltoall compared to scatter based alltoall implementations and MPI-based alltoall vs. CCL-Alltoall version as has been observed for strong scaling. E.g. we continue to see increased compute time due to communication overlap for MPI backend as shown in
Figure~\ref{fig:cc_brk_weak}. For the CCL backend the difference in compute time with and without overlap is almost negligible. Also, for the MLPerf config, we see, how cost of communication goes down at first (up to 8 ranks) and then increases again due to different characteristics of alltoall and allreduce. This can be further confirmed from a cost break-down for alltoall and allreduce depicted in Figure~\ref{fig:comm_brk_weak}.
Finally, we see the compute cost (even with blocking communication) for the MLPerf config increases slowly as we weak-scale, instead we expect it to remain constant. A more detailed look revealed that this additional cost comes from the current data loader design which always reads the data for full global minibatch on each rank and with weak scaling that cost steadily grows. For small and large configs, we use random dataset which does not account for time spent in data loader.% so we do see the same compute-cost behavior there.

\subsubsection{\textbf{Scaling on 8-Socket Shared Memory System}}
\begin{figure*}[t!]
\centering
\includegraphics[width=0.69\columnwidth]{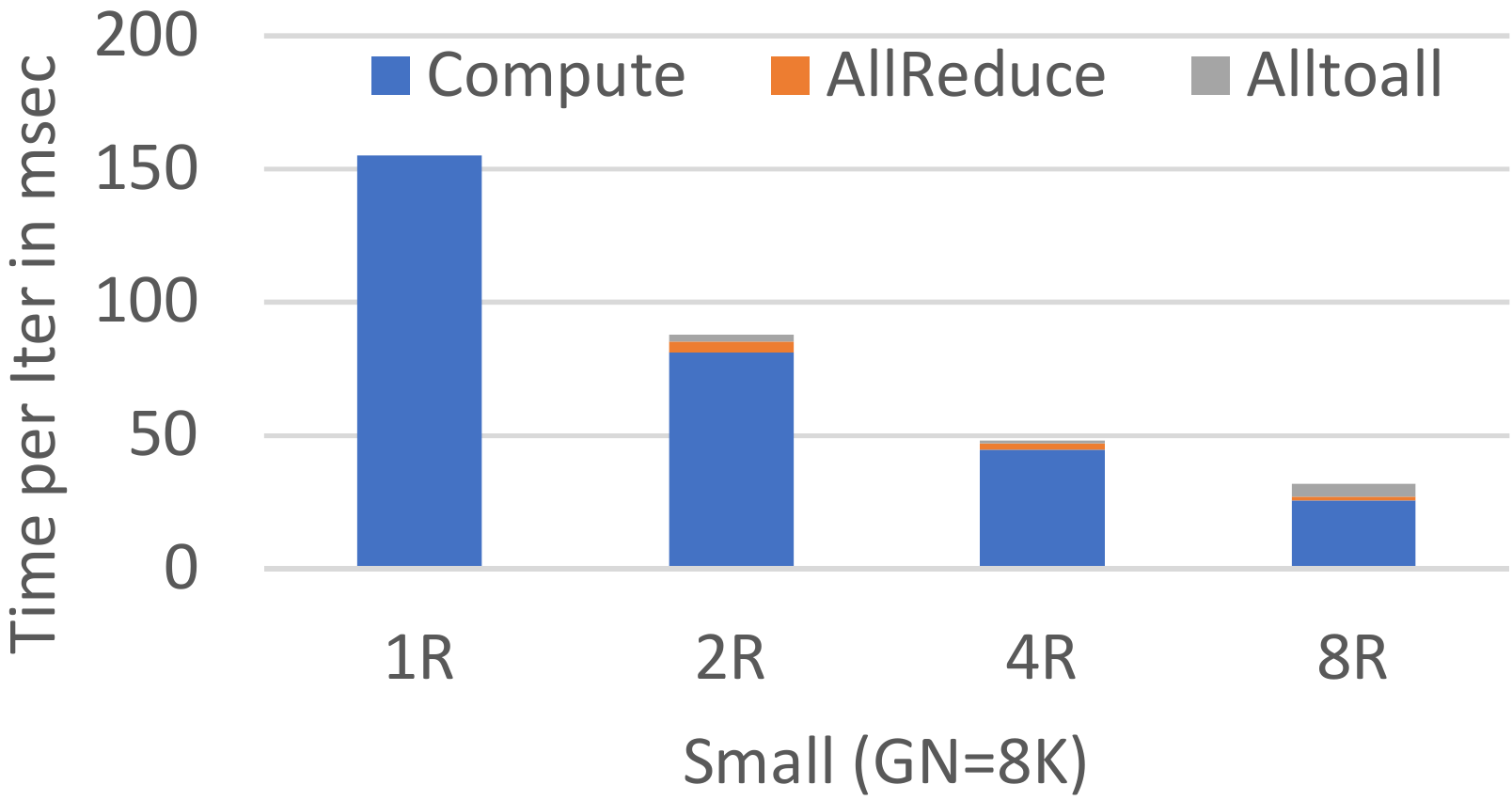}
\includegraphics[width=0.59\columnwidth]{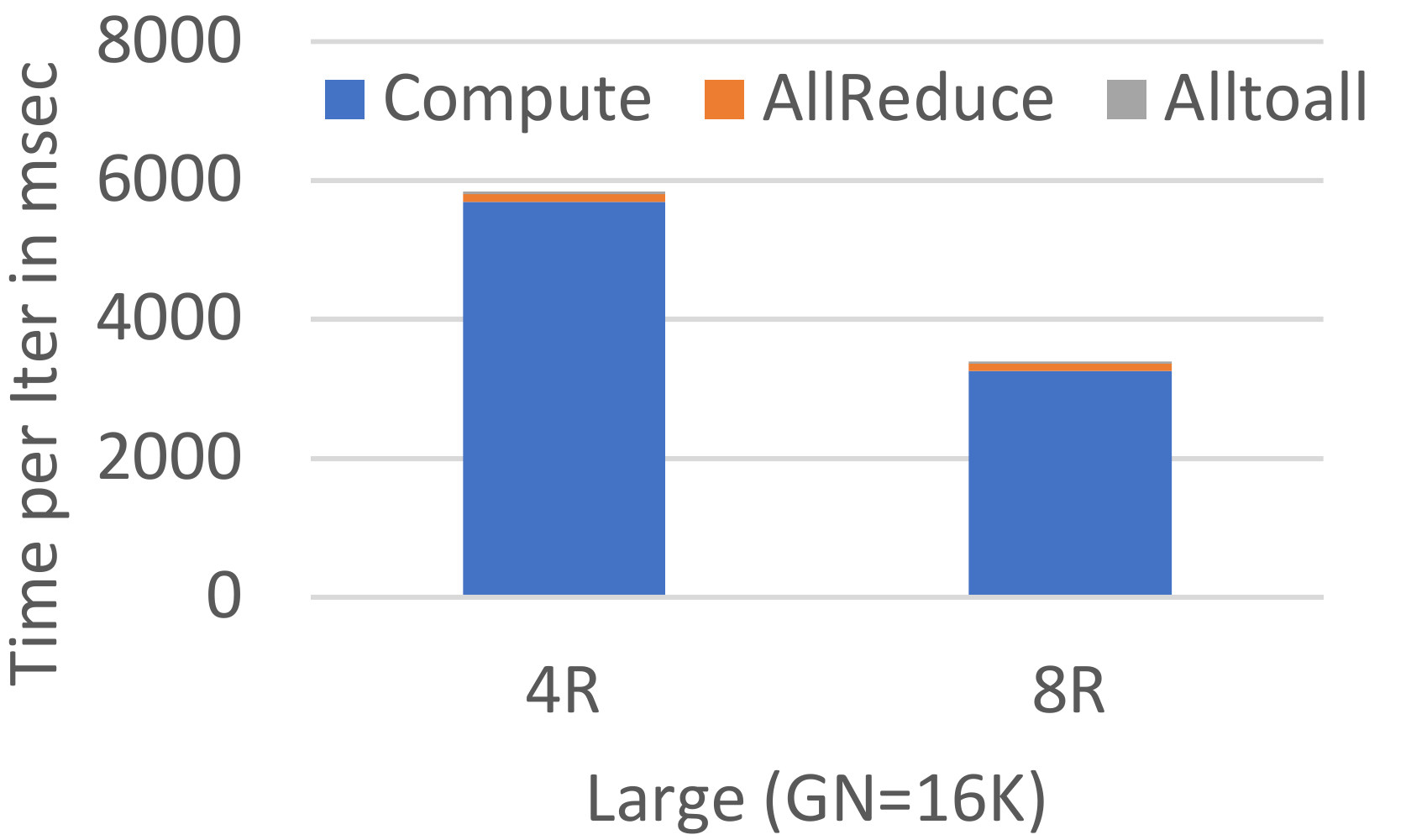}
\includegraphics[width=0.69\columnwidth]{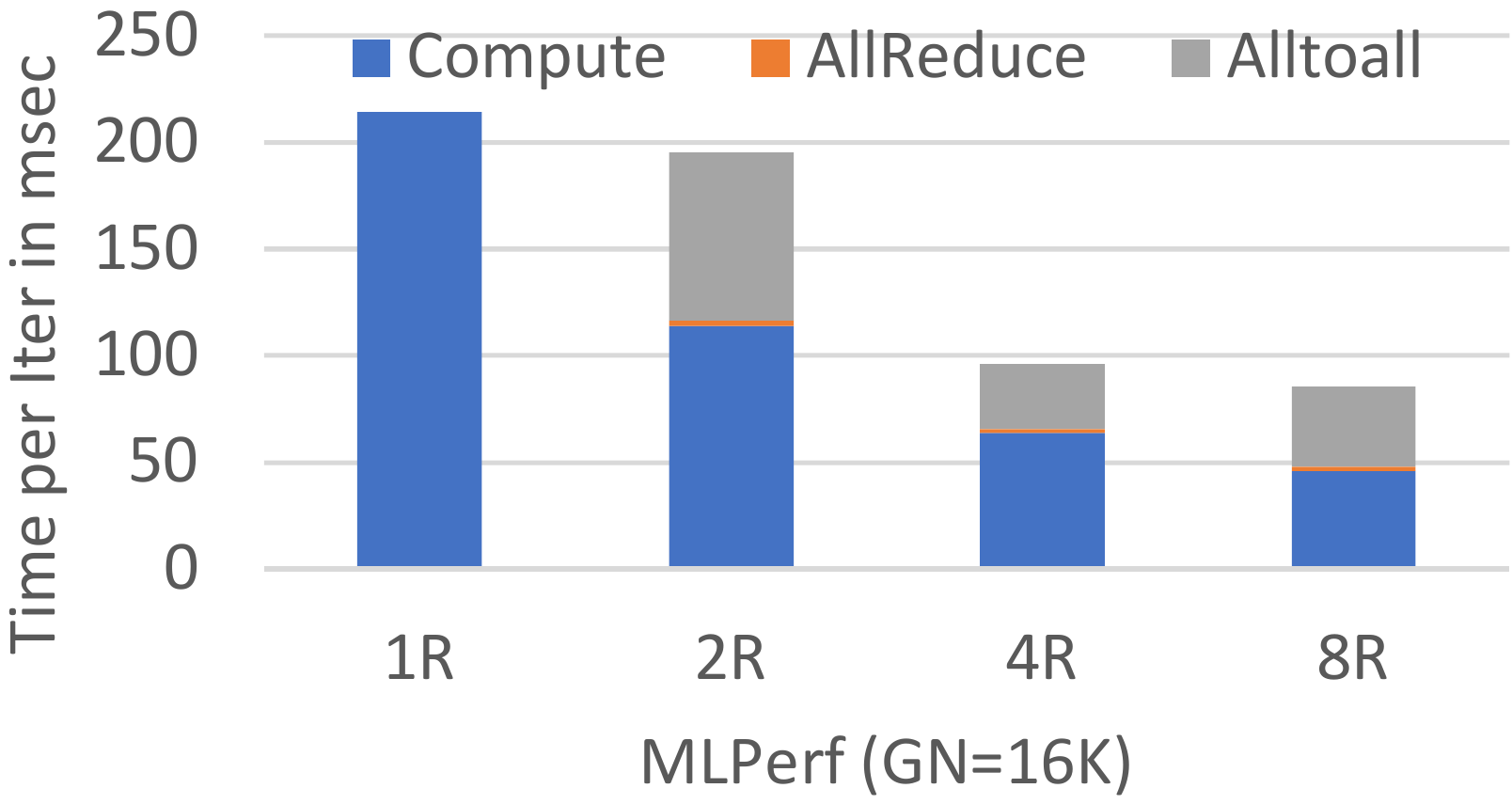}
\caption{Strong scaling performance on 8-socket shared-memory system}
\label{fig:kini_strong}
\end{figure*}

Finally, in Fig.~\ref{fig:kini_strong}, we discuss the strong scaling performance of the 8 socket shared-memory system which exhibits similar behavior to a cluster with similar node count. The only difference is, that the cost of alltoall does not decrease from 4 to 8 sockets as expected. This is because the alltoall implementation is not optimally tuned for twisted-hypercube connectivity on the system and so links are not utilized optimally. Additionally, even optimal algorithms would need multiple rounds of communication such that only 1.5x can be 
expected when comparing 4 to 8 nodes. This is evident for the MLperf config. Nevertheless,
as the 8 socket system fits all our workloads under investigation, it can be seen as a small cluster in an appliance form-factor as no external high-performance fabric is needed. So we can understand why Facebook prefers such a design in a loosely-coupled, Ethernet-based cloud datacenter~\cite{8327042,Park2018DeepLI}. 

\section{Split-SGD for native BFLOAT16 training}
\label{sec:splitsgd}

\begin{figure}[!t]
  \begin{center}
  \includegraphics[width=0.9\linewidth]{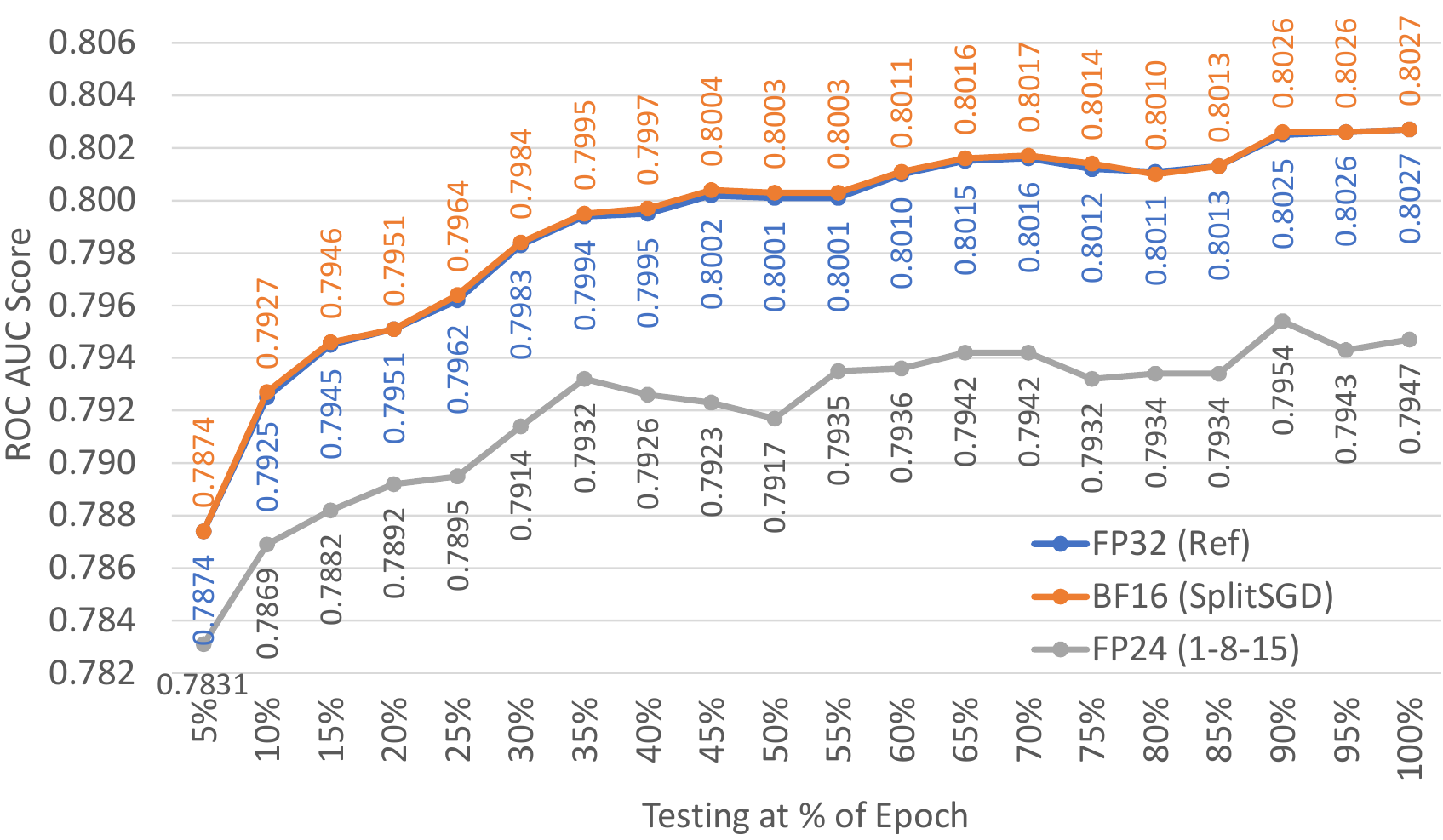}
  \end{center}
  \caption{Training accuracy with Mix precision BF16}
  \label{fig:bf16_cvg}
\end{figure}

BFLOAT16 is a new, but no-standard, floating-point format
\cite{google-bf16,intel-bf16} that is gaining
traction due to its ability to work well in machine learning algorithms, 
in particular deep learning training. In contrast to the IEEE754-standardized
16bit (FP16) variant, BFLOAT16 does not compromise at all on range when being
compared to FP32. As a reminder, FP32 numbers have 8 bits of exponent and 24
bits of mantissa (one implicit). BFLOAT16 cuts 16 bits from the 24-bit
FP32 mantissa to create a 16-bit floating point datatype. In contrast FP16,
roughly halves the FP32 mantissa to 10 explicit bits and has to reduce the
exponent to 5 bits to fit the 16-bit datatype envelope. BFLOAT16 therefore
perfectly aliases with the upper half of IEEE754-FP32 numbers.

Classic training approaches using BFLOAT16, FP16 or even
int16 require so-called master-weights, a full precision FP32 copy of the entire
model throughout the full training process~\cite{Kalamkar2019ASO,Micikevicius2018MixedPT,Das2018MixedPT}. With 16bit
''regular'' weights, this results into a 200\% (or 3X) overhead in storing the model.
As the model size in case of DLRM is mainly determined by the embedding tables, and
we are already starved for capacity, a mixed precision solver with FP16 would require 
hundreds of Gigabytes more capacity than a vanilla FP32 approach.

Instead, our Split-SGD-BF16 solver aims at efficiently exploiting the aforementioned
aliasing of BF16 and FP32. Therefore it reduces the overhead and being
equal to FP32 training, wrt. to capacity requirements, as master weights are implicitly 
stored. The trick is that we do not store FP32 values as a single tensor in a classic
fashion. Instead, we split them into their high and low 16bit parts. 
First we store all 16 MSBs of the FP32 numbers and then all 16 LSBs of the numbers as two separate tensors.
The 16 MSBs represent a valid BFLOAT16 number and we store it as part of model. We exclusively use
those in the forward and backward passes, whereas lower bits are only required in optimizer 
so we store them as additional state in optimizer and the actual update uses both, 
MSBs and LSBs and runs therefore a
fully FP32-accurate update. That means 66\% of the training passes enjoy a 2x bandwidth
reduction without negatively affecting the accuracy.
Figure~\ref{fig:bf16_cvg} demonstrates that this
technique allows us to train the DLRM MLPerf configuration to state-of-the-art with an error of 
less than 0.001\%. Please note, in this plot we also use a bit-accurate software 
emulation of Intel Xeon Cooper Lake BFLOAT16 dot-product instructions
\texttt{vdpbf16ps}~\cite{intel-sdm}.
When silicon is available,
this will help to also significantly speed-up the MLP portions as well. The described
concept of Split-SGD-BF16 is independent of the workload and can be transferred to any 
BFLOAT16 deep learning training task.

In addition, 
we tried to run with only 8 additional LSBs. However, these are not enough to train DLRM
to required accuracy. Note, that we tried to replicate FP16 embedding training with 
stochastic quantization as described in~\cite{Zhang2018TrainingWL}. Unfortunately, we 
were not able to train various DLRM configuration to state-of-the-art using SGD. We believe, 
more work is needed to generalize~\cite{Zhang2018TrainingWL} beyond the very
simple model problems presented.
\section{Related Work}
\label{sec:relwork}

While we believe this paper is the first of its kind
in covering large-scale DLRM training with an in depth analysis, 
some high level considerations for DLRM training were published in~\cite{Naumov2020DeepLT}. Apart from training, several studies have analyzed inference properties of DLRM~\cite{Gupta2019TheAI,Gupta2020DeepRecSysAS}
and this even pertains to FPGA acceleration~\cite{Centaur}. 
All these works have in common, that 
the CPU code, used as a baseline, is not as highly optimized as ours. 
With respect to 
large-scale training, results for CNNs have been shown here \cite{Dryden:2019:CFP:3295500.3356207,DBLP:journals/corr/abs-1711-04325,DBLP:journals/corr/abs-1709-05011,DBLP:journals/corr/abs-1811-05233} and RNN/LSTMs are covered in
\cite{8891019,You:2019:LTL:3295500.3356137}.
Large-scale HPC and Deep Learning on on-premises systems and in the cloud accelerated
by MPI progressions threads has been previously discussed in~\cite{7516082,10.1109/IPDPS.2014.113,Sridharan2018OnSD,4663774}.
The efficient implementation of various deep learning compute operations on Intel
CPUs such as convolutions and LSTM cells were 
discussed in~\cite{georganas2020ipdps,Georganas:2018:AHD:3291656.3291744,8891019}
and our work is based on these findings.
BFLOAT16 training with master weights on CPUs and without Split-SGD has been shown for a variety of workloads in~\cite{Kalamkar2019ASO}.
For completeness we have to note that NVIDIA also open-sourced a recommender system architecture 
which seems to be  more aligned with NCF and less complex 
than DRLM \cite{hugectr2019}. 

\section{Conclusion and Future Work}
\label{sec:conclusion}

In this paper we presented how programming methodologies and hardware architectures
from the field of HPC can be used to significantly speed-up training AI topologies, 
specifically the Deep Learning Recommender Models (DLRM). On Skylake/Cascade Lake 
CPUs we achieve a performance boost of more than two orders of magnitude (110$\times$)
on a single socket and studied strong and weak scaling with excellent results for
various problems sizes. This is true for large shared-memory nodes and clusters with
up to 64 CPU sockets. In addition to making the DLRM benchmark framework fit for
CPU cluster architectures of today, we also introduced and enabled DLRM with
a novel SGD optimizer. This leverages the BFLOAT16 datatype which is soon to be supported by various CPU architectures while matching FP32 accuracies. 
Last but not least we demonstrated that a single socket CPU is ~2x faster than a single V100 GPU, while we expect the GPU could be in theory 2x-3x faster if the code would be fully optimized. As the CPU cluster is not limited by capacity per socket, the best
sweet spot between MLP and embedding performance, or alltoall and allreduce, respectively, can be picked. 

\FloatBarrier
%\IEEEtriggeratref{19}

\bibliographystyle{IEEEtran}
%\clearpage
\bibliography{references}

%\scriptsize
%\noindent Optimization Notice: Software and workloads used in
%performance tests may have been optimized for performance only on
%Intel microprocessors.  Performance tests, such as SYSmark and
%MobileMark, are measured using specific computer systems,
%components, software, operations and functions.  Any change to any
%of those factors may cause the results to vary.  You should
%consult other information and performance tests to assist you in
%fully evaluating your contemplated purchases, including the
%performance of that product when combined with other products.
%For more information go to http://www.intel.com/performance.

%\noindent Intel, Xeon, and Intel Xeon Phi are trademarks of Intel Corporation in the %U.S. and/or other
%%countries.

%\normalsize

%\cleardoublepage
%\input{chapters/appendix.tex}

\end{document}